\documentclass[twocolumn]{aastex631}

\usepackage{CJKutf8}

\usepackage{amsmath}
\usepackage{units}
\usepackage{xcolor}
\usepackage{ulem}

\hypersetup{linkcolor=red, citecolor=blue, urlcolor=blue}

\renewcommand*{\emph}[1]{\textit{\textbf{#1}}}

\shorttitle{Surviving He-star companions of SNe Ia}
\shortauthors{Z.-W. Liu et al.}

\begin{document}

\title{Signatures of a surviving helium-star companion in Type Ia supernovae and constraints on the progenitor companion of SN~2011fe}

\correspondingauthor{Zheng-Wei Liu}
\email{zwliu@ynao.ac.cn}

\author[0000-0002-7909-4171]{Zheng-Wei Liu}
\affiliation{Yunnan Observatories, Chinese Academy of Sciences (CAS), 396 Yangfangwang, Guandu District, Kunming 650216, China}
\affiliation{Key Laboratory for the Structure and Evolution of Celestial Objects, CAS, Kunming 650216, China}
\affiliation{University of Chinese Academy of Science, Beijing 100012, China}

\author[0000-0002-4460-0097]{Friedrich K. R\"{o}pke}
\affiliation{Zentrum f{\"u}r Astronomie der Universit{\"a}t Heidelberg, Institut f{\"u}r Theoretische Astrophysik, Philosophenweg 12, 69120 Heidelberg, Germany}
\affiliation{Heidelberger Institut f{\"u}r Theoretische Studien, Schloss-Wolfsbrunnenweg 35, 69118 Heidelberg, Germany}

\author[0000-0003-1356-8642]{Yaotian Zeng}
\affiliation{Yunnan Observatories, Chinese Academy of Sciences (CAS), 396 Yangfangwang, Guandu District, Kunming 650216, China}
\affiliation{Key Laboratory for the Structure and Evolution of Celestial Objects, CAS, Kunming 650216, China}
\affiliation{University of Chinese Academy of Science, Beijing 100012, China}

\begin{abstract}

Single-degenerate (SD) binary systems composed of a white dwarf and a non-degenerate helium (He)-star companion have been proposed as the potential progenitors of Type Ia supernovae (SNe Ia).  The He-star companions are expected to survive the SN Ia explosion in this SD progenitor model.  In the present work, we map the surviving He-star companion models computed from our previous three-dimensional hydrodynamical simulations of ejecta-companion interaction into the one-dimensional stellar evolution code \texttt{Modules for Experiments in Stellar Astrophysics} (\texttt{MESA}) to follow their long-term evolution to make predictions on their post-impact observational properties, which can be helpful for searches of such surviving He-star companions in future observations.  By comparing with the very late-epoch light curve of the best observed SN Ia, SN~2011fe, we find that our surviving He-star companions become significantly more luminous than SN~2011fe about $1000\,\mathrm{d}$ after the maximum light.  This suggests that a He star is very unlikely to be a companion to the progenitor of SN~2011fe.

\end{abstract}


\section{Introduction} 
\label{sec:introduction}

Type Ia supernovae (SNe Ia) have been used as cosmic distance indicators because their peak luminosities can be empirically standardized by the so-called `Phillips relation' \citep{Phillips1993, Phillips1999}, which has led to the discovery of the accelerating expansion of the Universe \citep{Riess1998,Schmidt1998,Perlmutter1999}. In addition, SNe Ia play a fundamental role in placing constraints on the equation of state of dark energy. It is generally thought that SNe Ia result from thermonuclear explosions of white dwarfs in binary systems \citep{Hoyle1960}. However, it is still hard to reach a consensus on the fundamental aspects of the nature of SN Ia progenitors and their actual explosion mechanism from both, the theoretical and observational side \citep[e.g.,][]{Hillebrandt2013, Maoz2014,Soker2019}.  Over the past decades, a range of models has been proposed to cause the explosion of white dwarfs (WDs) giving rise to SNe Ia. The main questions are how the star reaches the explosive conditions, i.e., the progenitor channel, and how the explosion proceeds, i.e., the explosion scenario. For the first, the main distinction is into the single degenerate (SD; e.g., \citealt[][]{Whelan1973, Nomoto1982,Hachisu1996,Han2004}) and the double degenerate (DD; e.g., \citealt[][]{Iben1984, Webbink1984, Dan2011}) channels. The explosion mechanism depends mainly on the question of whether the WD explodes near the Chandrasekhar mass \citep[e.g.,][]{Nomoto1984,Roepke2007a,Roepke2007b,Seitenzahl2013,Jordan2012,Fink2014,Lach2021,Lach2021b,Livio2003,Ilkov2012} or at a mass below this limit \citep[e.g.,][]{Woosley1986, Fink2007, Shen2007, Sim2010, Pakmor2010, Townsley2019, Gronow2020, Gronow2021,Benz1989, Rosswog2009,Kushnir2013}. To provide important clues on the yet poorly understood origin and explosion mechanism of SNe Ia, one needs to compare the observational features predicted by different progenitor models with the observations.

In the SD scenario, the WD accretes matter from a non-degenerate companion star through Roche-lobe overflow to trigger a thermonuclear explosion when its mass approaches the Chandrasekhar-mass limit \citep[e.g.,][]{Whelan1973, Han2004,Liu2018,Liu2020aa}, in which the donor star could be either a  main-sequence (MS), a slightly evolved MS, a red-giant (RG), or a helium (He) star.  

On the one hand, a non-degenerate companion star is much brighter than a WD; a luminous source (i.e. the non-degenerate companion) is therefore expected to be detected in pre-explosion image at position of the SN Ia. Analyzing pre-explosion images at the SN position provides a direct way to identify the SD progenitor scenario \citep[][]{Foley2014, McCully2014}. The pre-explosion observable properties of different non-degenerate donors at the moment of SN Ia explosion in the SD scenario have been comprehensively addressed for normal SNe Ia \citep{Han2008} and SNe Iax\footnote{SNe Iax potentially form the most common subclass of SNe Ia, with an estimated rate of occurrence of about 5\%--30\% of the total SN Ia rate \citep{Foley2013}. Recent work suggests that weak deflagrations of a Chandrasekhar-mass WD is able to reproduce the characteristic observational features of bright SNe Iax \citep[e.g.,][]{Jordan2012,Kromer2013,Lach2021}.} \citep{Liu2015}. To date, no companion star of a normal SNe Ia has been directly confirmed in pre-explosion images \citep[e.g.][]{Maoz2008, Li2011, Kelly2014}. However, a blue luminous source has been detected in pre-explosion image of an SN Iax event, SN~2012Z \citep{McCully2014}. This pre-explosion luminous source (i.e. SN~2012Z-S1) has been interpreted as a He-star companion  to the exploding WD \citep[][]{McCully2014, Liu2015}. Interestingly, late-time observations taken about 1400 days after the explosion by the Hubble Space Telescope have shown that SN 2012Z is brighter than the normal SN~2011fe by a factor of two at this epoch \citep{McCully2021}. This excess flux is suggested to be a composite of several sources: the shock-heated companion, a bound remnant that  could drive a wind, and light from the SN ejecta due to radioactive decay \citep{McCully2021}.

On the other hand, non-degenerate companion stars are expected to survive a SN Ia explosion in the SD scenario \citep[e.g.][]{Wheeler1975, Marietta2000, Pakmor2008, Liu2012,Liu2013c,Liu2013a, Liu2013b,Liu2021b,Pan2012a,Boehner2017,Bauer2019,Zeng2020}. Searching for the surviving companion star in nearby SN remnants (SNRs) has been considered a promising way to test the SD progenitor scenario \citep[e.g.][]{Kerzendorf2009, Ruiz-Lapuente2019}. In the past few decades, many studies have been focused on searching for the surviving companion star predicted by the SD scenario in the Galactic supernova remnants (SNRs) Tycho \citep[e.g.,][]{Ruiz-Lapuente2004,Ruiz-Lapuente2019,Fuhrmann2005,Ihara2007,Gonzalez2009,Kerzendorf2009,Kerzendorf2013,Kerzendorf2018a,Bedin2014}, Kepler \citep[e.g.,][]{Kerzendorf2014,Ruiz-Lapuente2018}, and SN~1006 \citep[e.g.,][]{Gonzalez2012,Kerzendorf2012,Kerzendorf2018b}, and in some SNRs in the Large Magellanic Cloud (LMC): SNR~0509–67.5 \citep[e.g.,][]{Schaefer2012,Pagnotta2014,Litke2017}, SNR~0519–69.0 \citep[e.g.,][]{Edwards2012,Li2019}, SNR~0505–67.9 (DEML71, \citealt{Pagnotta2015}), SNR~0509–68.7 (N103B, e.g., \citealt[][]{Pagnotta2015,Li2017}), and SNR~0548–70.4 \citep[e.g.,][]{Li2019}.  To date, no surviving companion could be firmly identified in these SNRs. \citet{Shen2018}  suggest that three hypervelocity runaway stars detected in \textit{Gaia}'s second data release \citep[][]{Gaia2016,Gaia2018} are WD companions that survived so-called dynamically driven double-degenerate double-detonation (`D6') SNe Ia.

\begin{figure*}
  \begin{center}
    {\includegraphics[width=0.96\textwidth, angle=360]{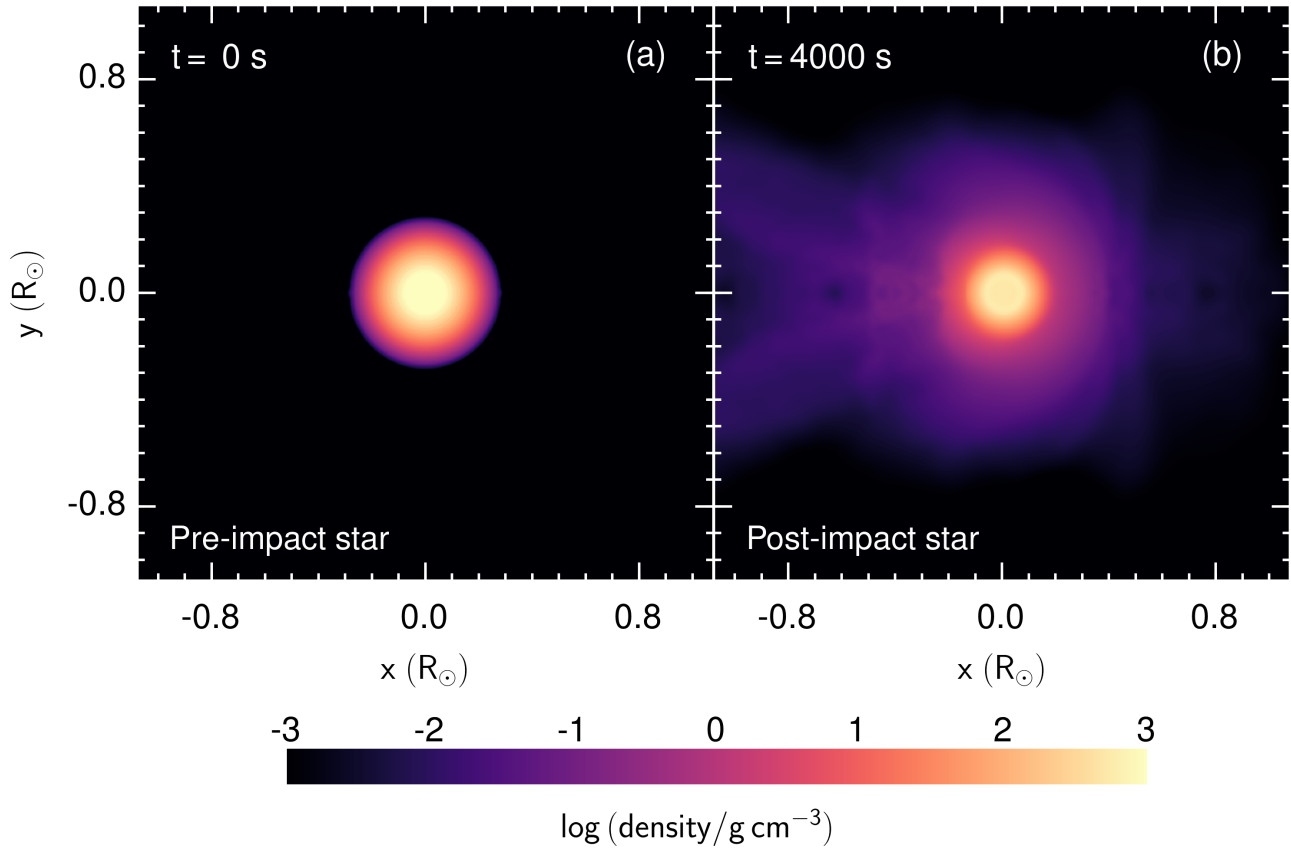}}
    \caption{He-star companion at the pre-impact phase (\textsl{left panel}) and at the end of 3D hydrodynamical simulations (\textsl{right panel}) of ejecta-donor interaction in \citet[][]{Liu2013b}. The panels show density slices in the orbital plane. Here, only the so-called  `He01' is shown as an example. The color scale gives the logarithm of the mass density in $\rm{g\,cm^{-3}}$.}
\label{Fig:he}
  \end{center}
\end{figure*}

The surviving companion stars are expected to present distinct observational signatures after the explosion due to the mass-stripping, shock heating and enrichment with heavy elements from SN ejecta during the ejecta-companion interaction. Using one-dimensional (1D) stellar evolution codes, some previous studies have investigated the post-impact properties of surviving companions of SNe Ia. They followed the long-term evolution of the companion models that either were produced from three-dimensional (3D) hydrodynamical simulations of ejecta-companion interaction \citep[e.g.,][]{Pan2012b,Pan2013,Liu2021a,Liu2021b} or have mimicked the interaction with the supernova eject by a rapid mass loss and extra heating \citep[e.g.,][]{Podsiadlowski2003,Shappee2013}. All previous studies suggest that the surviving companion stars of SNe Ia generally become significantly more luminous after the impact during the thermal re-equilibration phase.

\begin{table*}\renewcommand{\arraystretch}{1.3}
\fontsize{9}{11}\selectfont
\begin{center}
\caption{Helium-star companion models and results of this work.} \label{table:1}
\centering
\begin{tabular}{lccccccccccc}    
\hline\hline
 Name & $M^{\mathrm{\,f}}_{\mathrm{2}}$   &$\mathrm{log}\,P^{\mathrm{f}}$   & $R^{\mathrm{f}}_{\mathrm{2}}$ & $V^{\mathrm{\,f}}_{\mathrm{orb}}$ & $V^{\mathrm{\,f}}_{\mathrm{rot}}$ & $\mathrm{log}\,T^{\mathrm{f}}_{\mathrm{eff}}$   & $\Delta M/M^{\mathrm{\,f}}_{2}$ & \ \ $V_{\rm{kick}}$    & $E_{\mathrm{in}}$ & $t_{\mathrm{peak}}$ & $L_{\mathrm{peak}}$\\
    &($M_{\rm{\odot}}$)                   & (days)     & ($10^{10}\,\mathrm{cm}$) & ($\mathrm{km\,s^{-1}}$) & ($\mathrm{km\,s^{-1}}$) & ($K$)&  (\%) & ($\mathrm{km\,s^{-1}}$) &($10^{49}\,\mathrm{erg}$) & ($\mathrm{yr}$) & ($10^{4}\,L_{\odot}$)\\

\hline
   He01 & 1.24   & -1.34 &  1.91 & 432 & 301 & 4.70 & 2.2 & 66 & 1.68 & 14.94 & 1.65\\
   He02 & 1.01   & -1.12 &  2.48 & 387 & 237 & 4.85 & 5.6 & 59 & 1.19 & 29.75 & 0.58\\
   He01r & 1.24  & -1.34 &  1.91 & 432 & 301 & 4.70 & 2.3 & 67 & 1.72 & 22.50 & 1.53\\
   He02r & 1.01  & -1.12 &  2.48 & 387 & 237 & 4.85 & 5.7 & 60 & 1.19 & 72.36 & 0.44\\
\hline
   0p8\_He01r & 1.24  & -1.34 &  1.91 & 432 & 301 & 4.70 & 1.5 & 40 & 1.47 & 8.05  & 1.95\\
   1p0\_He01r & 1.24  & -1.34 &  1.91 & 432 & 301 & 4.70 & 1.9 & 53 & 1.58 & 15.76 & 1.72\\
   1p4\_He01r & 1.24  & -1.34 &  1.91 & 432 & 301 & 4.70 & 2.7 & 76 & 1.79 & 34.18 & 1.60\\
   1p6\_He01r & 1.24  & -1.34 &  1.91 & 432 & 301 & 4.70 & 3.0 & 86 & 1.88 & 42.24 & 1.45\\
\hline
\end{tabular}
\end{center}

\textbf{Note.} Here, the superscript letter `f' means the moment of SN Ia explosion. $M^{\mathrm{\,f}}_{\mathrm{2}}$, $P^{\mathrm{f}}$, $R^{\mathrm{f}}_{\mathrm{2}}$, $V^{\mathrm{\,f}}_{\mathrm{orb}}$, $V^{\mathrm{\,f}}_{\mathrm{rot}}$ and $\mathrm{log}\,T^{\mathrm{f}}_{\mathrm{eff}}$ denote the final mass, orbital period, radius, orbital velocity, rotational velocity and effective temperature of the donor star, respectively; $\Delta M$ and $V_{\rm{kick}}$ are the total amount of stripped donor mass and the kick velocity received by the donor star due to the ejecta-donor interaction; $E_{\mathrm{in}}$ corresponds to the total amount of energy absorbed by the donor star during the interaction. `He01' and `He02' are two He-star companion models in \citet{Liu2013b}. A suffix `r' means that the orbital motion and spin of the He-star companion are included into 3D impact simulations. `0p8', `1p0', `1p4' and `1p6' mean that different kinetic energies of SN Ia ejecta of $0.8,1.0,1.4$ and $1.6\times10^{51}\,\mathrm{erg}$ were used for 3D impact simulations \citep[][see their Sec.~4.3.3]{Liu2013b}.\\

\end{table*}

\begin{figure*}
  \begin{center}
    {\includegraphics[width=0.48\textwidth, angle=360]{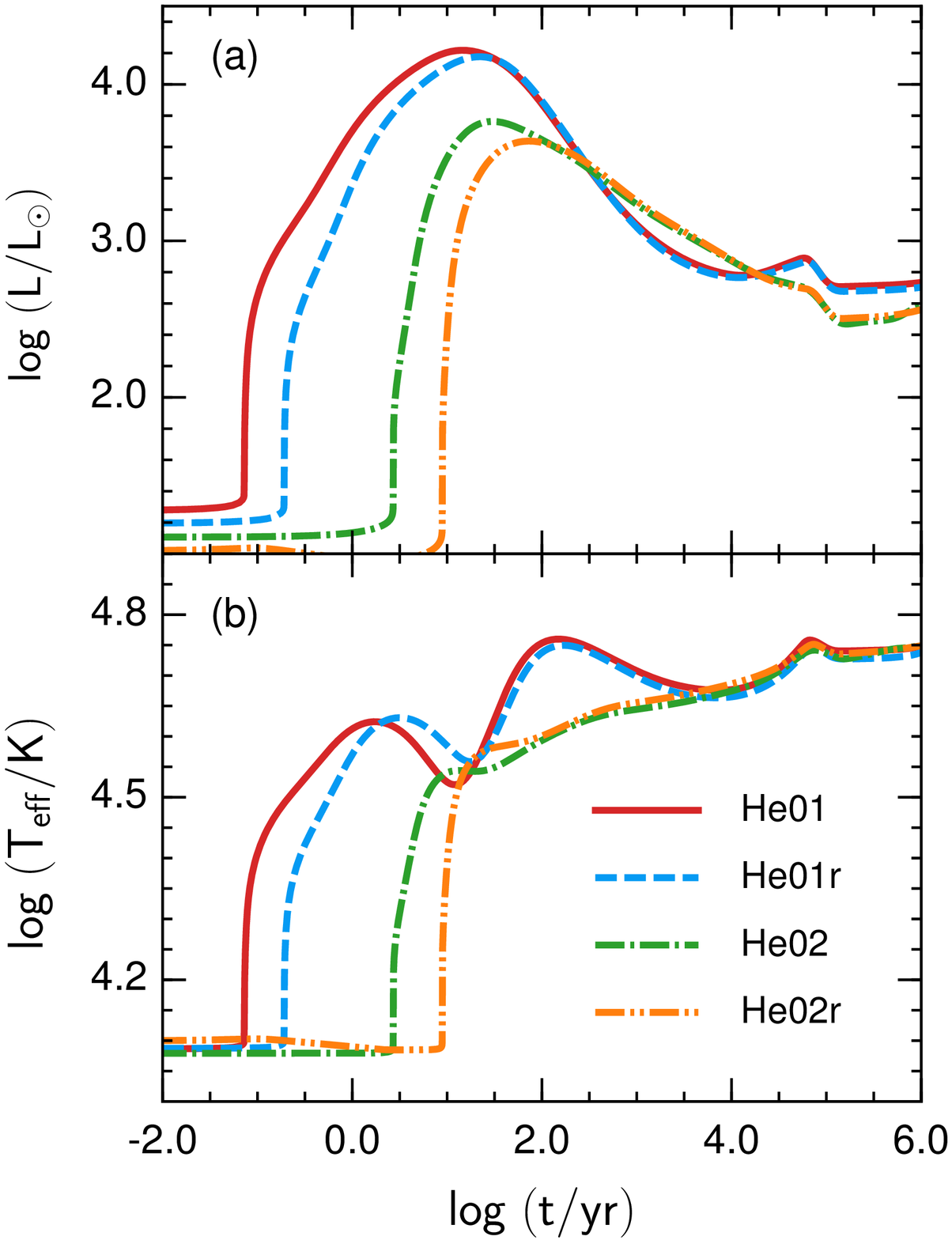}}
    \hspace{0.2in}
    {\includegraphics[width=0.48\textwidth, angle=360]{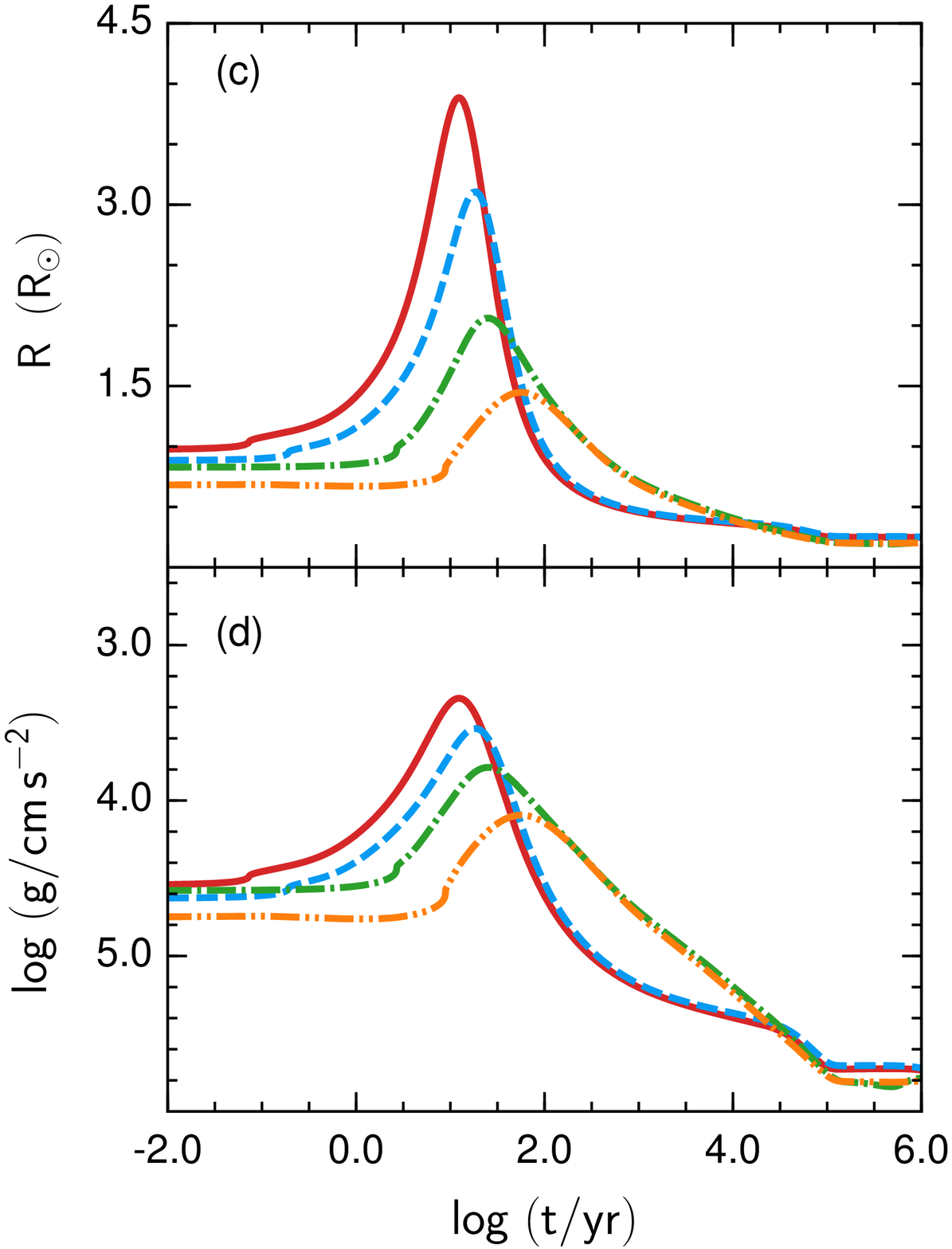}}
    \caption{Post-impact evolution of the photosphere luminosity $L$, effective temperature $T_{\mathrm{eff}}$, radius $R$, and surface gravity $g$ of surviving He star companions as functions of time. The red solid and green dash-dotted curves correspond to models He01 and He02 of \citet{Liu2013b} that do not include rotation. In contrast, models He01r and He02r were set up from our 3D impact simulations including binary orbital motion and spin.}
\label{Fig:2}
  \end{center}
\end{figure*}

\begin{figure*}
  \begin{center}
    {\includegraphics[width=0.48\textwidth, angle=360]{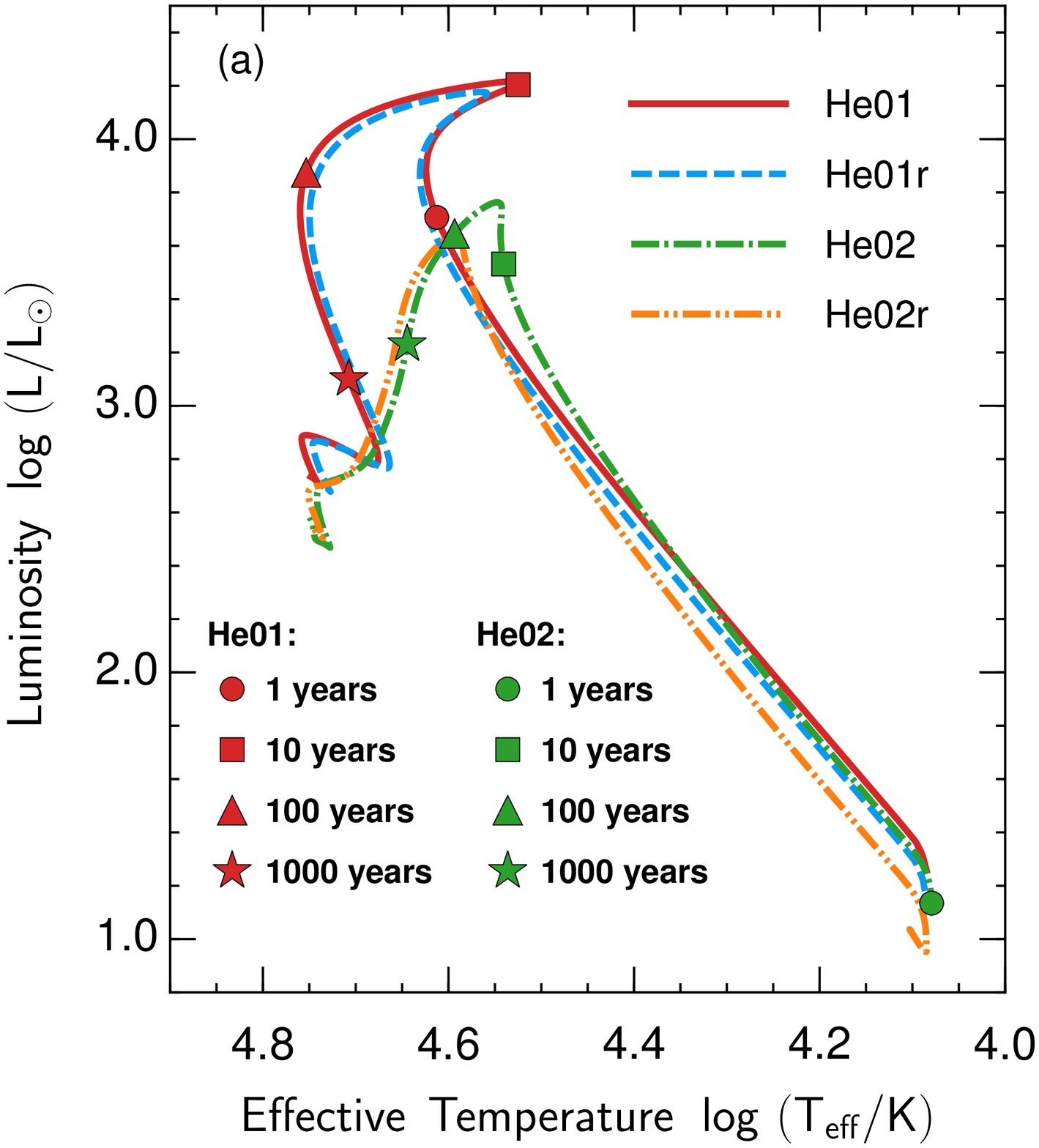}}
    \hspace{0.2in}
    {\includegraphics[width=0.48\textwidth, angle=360]{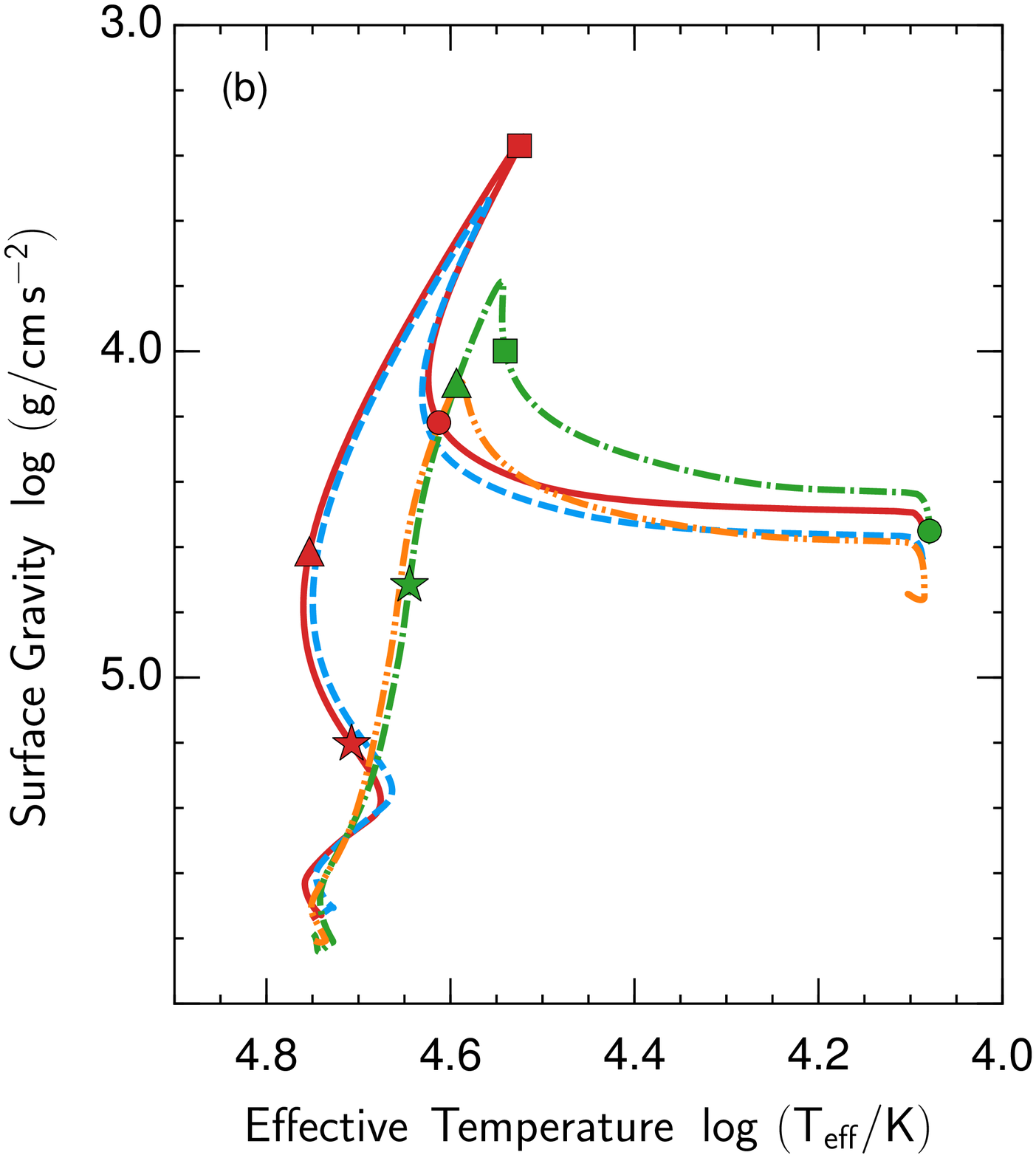}}
    \caption{Post-impact evolutionary tracks of two surviving He-star companion models (i.e., He01 and He02 model of \citealt{Liu2013b}) in the Hertzsprung-Russell Diagram (\textsl{Left Panel}) and surface gravity vs.\ temperature diagram (\textsl{Right Panel}). For a comparison, the results based on the impact simulations that do include the binary orbital motion and stellar spin  (i.e., the so-called `He01r' and `He02r' model) are given as a \textsl{blue dashes line} and  \textsl{yellow double-dotted line}. The \textsl{filled circle}, \textsl{square}, \textsl{triangle}, and \textsl{star markers} on the tracks present post-impact evolutionary phases of $1\,\mathrm{yr}$, $10\,\mathrm{yr}$, $100\,\mathrm{yr}$, and $1000\,\mathrm{yr}$ after the SN impact, respectively. Here, we have only marked different evolutionary phases of non-rotating models (i.e., the He01 and He02 model) because no significantly differences are observed in rotating models.}
\label{Fig:3}
  \end{center}
\end{figure*}

\begin{figure}
  \begin{center}
    {\includegraphics[width=0.49\textwidth, angle=360]{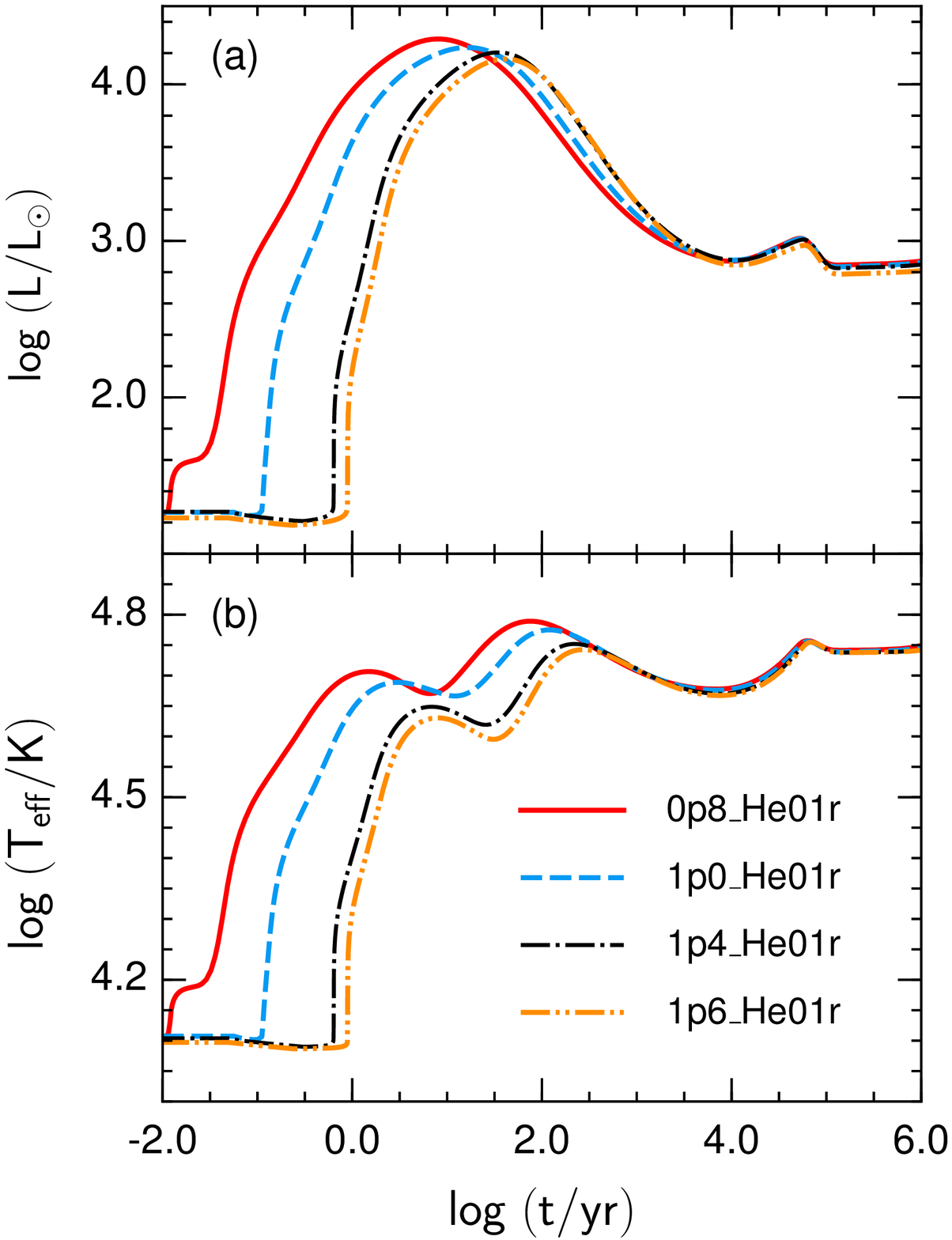}}
    \caption{Similar to \textsl{left-hand panel} of Fig.~\ref{Fig:3}, but for the cases artificially adopting different explosion energies (i.e., $E_{\mathrm{kin}}^{\mathrm{SN}}=0.8$, $1.0$, $1.4$ and $1.6\times10^{51}\,\mathrm{erg}$) in impact simulations of  \citet[][see their Sec.~4.3.3]{Liu2013b} for the He01r model.}
\label{Fig:4}
  \end{center}
\end{figure}

In our previous studies, we have performed 3D hydrodynamical simulations of ejecta-companion interaction within the He-star donor progenitor scenario of SNe Ia \citep{Liu2013b}, in which the smoothed-particle hydrodynamics (SPH) code \textsc{Stellar Gadget} \citep{Springel2001, Pakmor2012a} was adopted for our impact simulations, and a near-Chandrasekhar-mass explosion model of \citet[][]{Nomoto1984} (i.e., the so-called \textsc{W7} model) was used for representing normal SNe Ia. However, we did not address the long-term evolution and appearance of our surviving He-star companions. The main goal of this work is to provide observational properties of these surviving He-star companions by following their long-term post-impact evolution. Furthermore, we compare our results with the late-time observations of the best observed SN Ia SN~2011fe \citep[e.g.,][]{Shappee2017,Dimitriadis2017,Kerzendorf2017} to place constraints on its progenitor model.

As mentioned above, \citet{Pan2013} also explored the long-term evolution of surviving He-star companions of SNe Ia. However, the companion models used in their 3D hydrodynamic impact simulations were constructed by artificially adopting a constant mass-loss rate instead of a detailed binary evolution calculation to mimic the detailed binary evolutionary models. Our initial He-star companion stars (i.e., `He01' and `He02' model, see Table~\ref{table:1}) were followed through the full binary evolution \citep{Liu2013b}. In model He01, the He-star companion remains in its MS phase in He01 model until the onset of the SN explosion. In model He02, in contrast, it has evolved slightly into the subgiant
phase (i.e., the central He is exhausted). The detailed binary evolution \citep[][see their Sect.~2]{Liu2013b} and population synthesis calculations  \citep[][see their Figs.~3--5]{Wang2009} suggest that these two models are likely to represent progenitors of SNe Ia. To determine observational properties for searches for surviving companions, it is important to follow the long-term evolution of our models. In addition, only few surviving companion models have been explored by previous works.  An improvement to predictions on the observable features of surviving companions of SNe Ia requires further studies that cover a wider range of companion models.

\begin{figure}
  \begin{center}
    {\includegraphics[width=0.49\textwidth, angle=360]{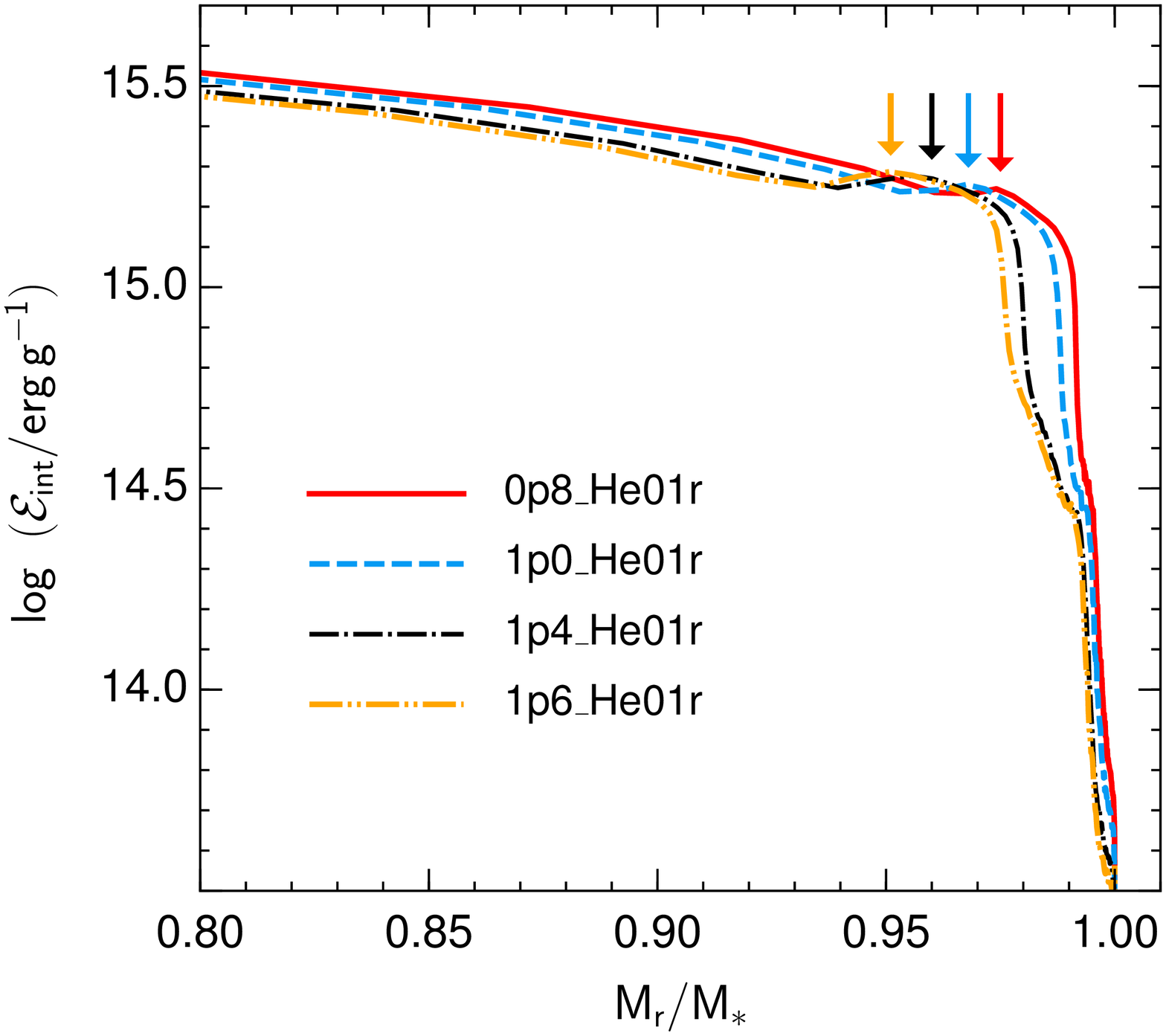}}
    \caption{Specific internal energy ($\mathcal{E}_{\mathrm{int}}$) of the surviving companion as a function of the mass fraction for the He01r model with different explosion energies (similar to Fig.~\ref{Fig:4}). For better visibility, only the outermost layers (i.e., $M_{\mathrm{r}}/M_{\ast}=0.8$--$1.0$) of the star are plotted. The local maxima indicated by arrows approximately correspond to the depths of energy deposition of different cases.}
\label{Fig:5}
  \end{center}
\end{figure}

\begin{figure}
  \begin{center}
    {\includegraphics[width=0.49\textwidth, angle=360]{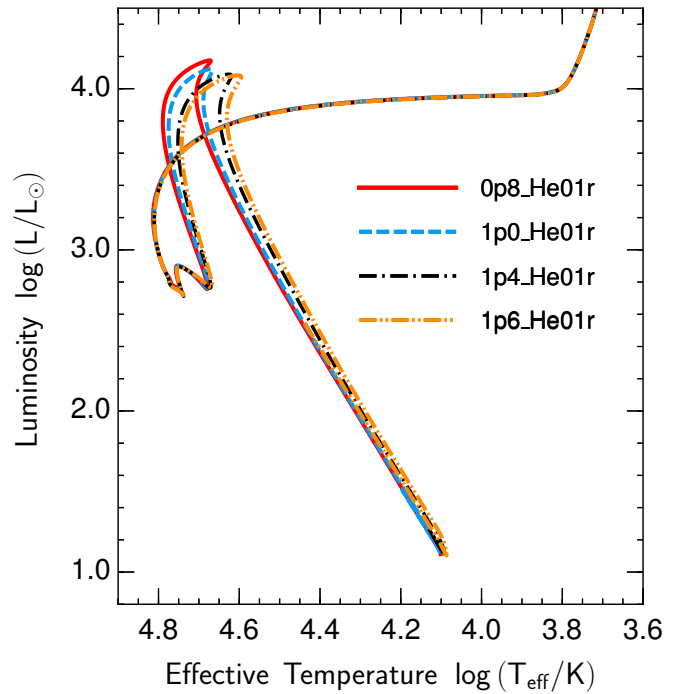}}
    \caption{Similar to Fig.~\ref{Fig:3}, but for the cases by artificially adopting different explosion energies in impact simulations of \citet{Liu2013b} for the He01r model.}
\label{Fig:6}
  \end{center}
\end{figure}

\begin{figure}
  \begin{center}
    {\includegraphics[width=0.49\textwidth, angle=360]{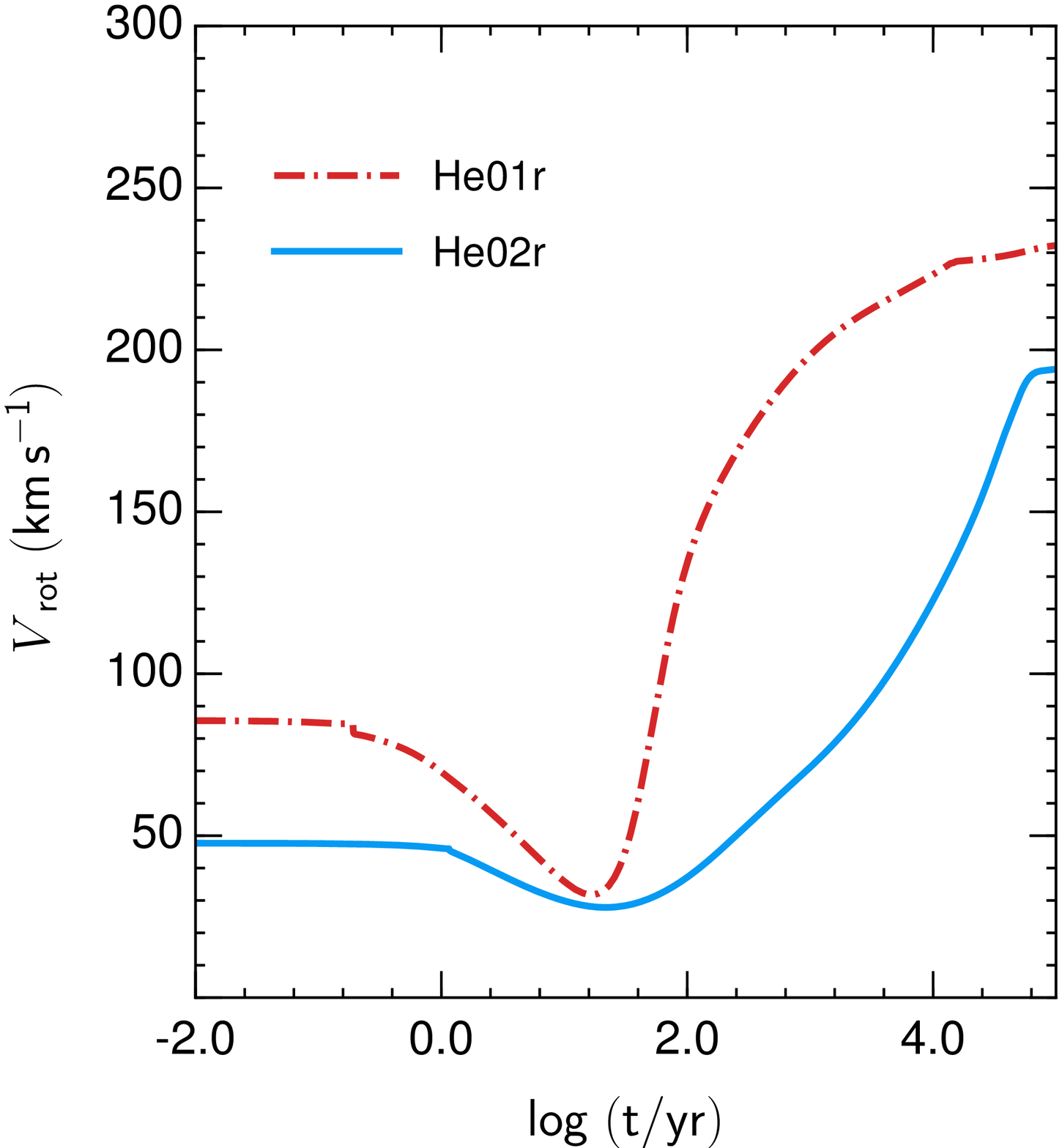}}
    \caption{Similar to Fig.~\ref{Fig:2}, but for the post-impact evolution of surface rotational speed of the star in model He01r (dash-dotted line) and He02r (solid line).}
\label{Fig:vrot}
  \end{center}
\end{figure}

\section{Mapping from 3D to 1D}

To provide post-impact observational properties of surviving companion stars for their identifications in historical SNRs, we use the 1D stellar evolution code \texttt{Modules for Experiments in Stellar Astrophysics} \citep[\texttt{MESA};][]{Paxton2011, Paxton2018} to follow the post-impact evolution of these stars for a long time up to the age of historical SNRs, i.e., a few times $100$--$1000\,\mathrm{yr}$. We use the methods described in \citet[][]{Liu2021a} and \citet[][]{Liu2021b} to construct suitable starting models for \texttt{MESA} based on 3D surviving He-star companion models (Fig.~\ref{Fig:he}) computed from our previous 3D hydrodynamical impact simulations \citep{Liu2013b}. For our rotating models, we also include the angle-averaged, radial angular momentum profile derived from the SPH output when constructing the initial models for the subsequent \texttt{MESA} calculations. We directly apply the relaxation routines of \texttt{MESA} to the outcomes of 3D SPH impact simulations to obtain the 1D stellar model with a chemical, angular momentum (which is set to be zero for non-rotating models) and thermal structure that closely matches that of the 3D companion remnant model. A detailed description of the use of the relaxation routines in \texttt{MESA} can be found in Appendix~B of \citet{Paxton2018}.  To test the validity of using \texttt{MESA} for modeling the long-term evolution of our surviving companion star models computed from 3D hydrodynamical impact simulations, we have compared the results of \texttt{MESA} calculations to those of the 1D hydrodynamic stellar evolution code \texttt{Kepler} \citep[][]{Weaver1978, Rauscher2002, Woosley2002,Heger2010} in previous work \citep{Liu2021b}. We find no significant difference between MESA and Kepler results for a given companion star model \citep[][]{Liu2021b}. Therefore, we only use \texttt{MESA} models here.

\section{Post-impact evolution of surviving companions}\label{sec:ms}

In this section, we present the numerical results of our 1D post-impact \texttt{MESA} calculations for eight surviving He-star companion models (Table~\ref{table:1}), for which the starting models -- after mapping from the 3D impact simulations --  have been relaxed with the routines provided in \texttt{MESA}. Following \citet[][]{Pan2013}, the models were ran with a fixed time-step of $10^{-8}\,\mathrm{yr}$ for the first $10^{-6}\,\mathrm{yr}$. Thereafter, the time-step was automatically determined in \texttt{MESA}. We also explore the dependence of post-impact evolution of the surviving He-star companion on the SN Ia explosion energy and their post-impact rotation properties. The initial parameters of all models and numerical results are also summarized in Table~\ref{table:1}.

\subsection{Post-impact properties}
\label{sec:he}

Figure~\ref{Fig:he} shows the evolution of density in the surviving companion during our impact simulations for model `He01' (Table~\ref{table:1}). The ejecta-companion interaction removes some He-rich material from the companion's surface and causes an energy deposition in the star  due to the SN impact and shock heating. We find that an energy deposition of $E_{\rm{in}}=1.68\times10^{49}\,\mathrm{erg}$ and $1.19\times10^{49}\,\mathrm{erg}$  for model He01 and He02, respectively. As a consequence, the star inflates after the explosion as shown in the right-hand panel of Fig.~\ref{Fig:he}. This leads to an increase in radius by a factor of two at the end of impact simulation (about $4000\,\mathrm{s}$ after the explosion). Here, we use the method of \citet[][see their Sect.~4.1]{Pan2013} to calculate the amount of energy deposition ($E_{\rm{in}}$)  by tracing the increase of binding energy of the remnant after SN impact.

\begin{figure*}
  \begin{center}
    {\includegraphics[width=0.48\textwidth, angle=360]{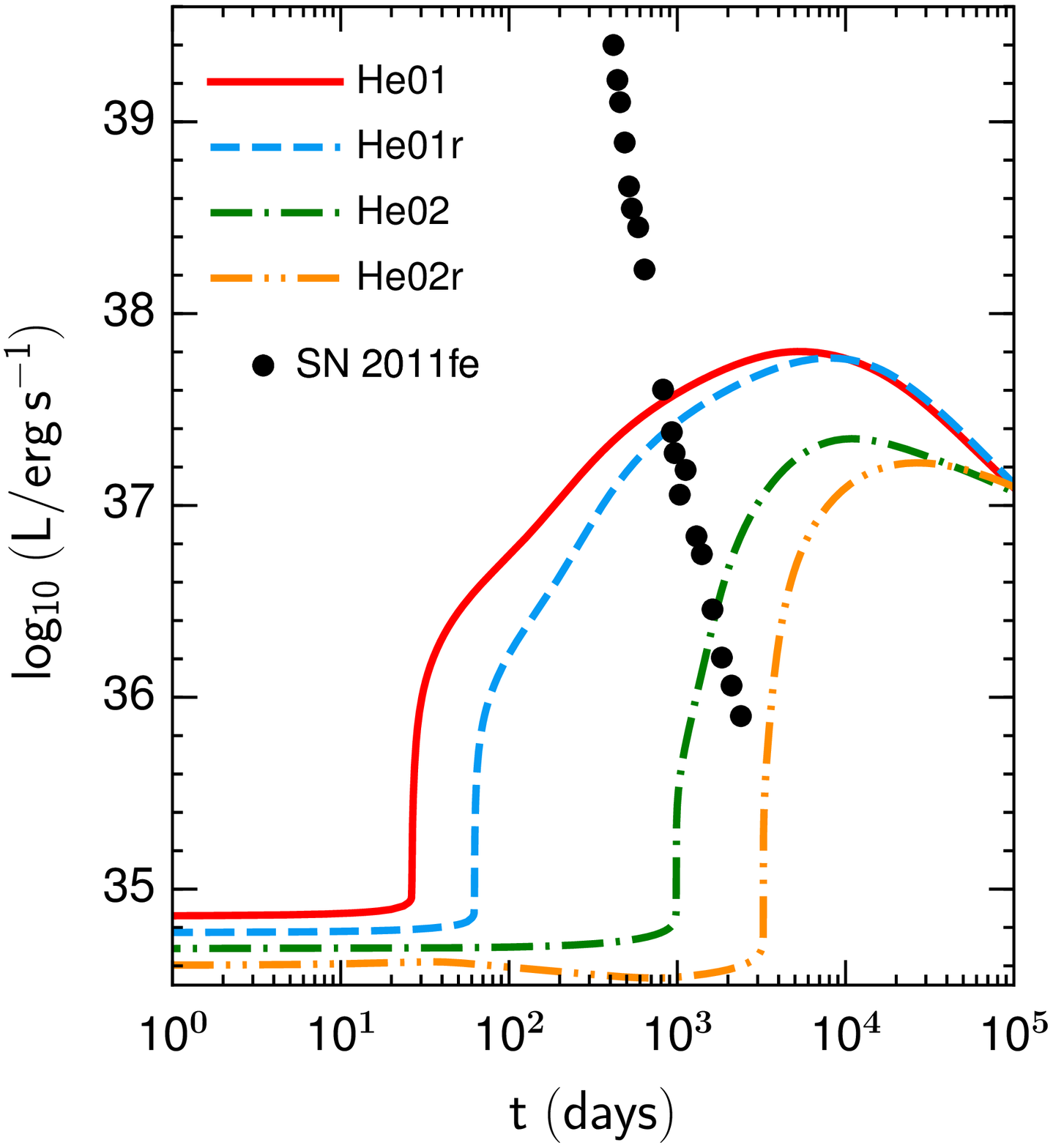}}
    \hfill
    {\includegraphics[width=0.48\textwidth, angle=360]{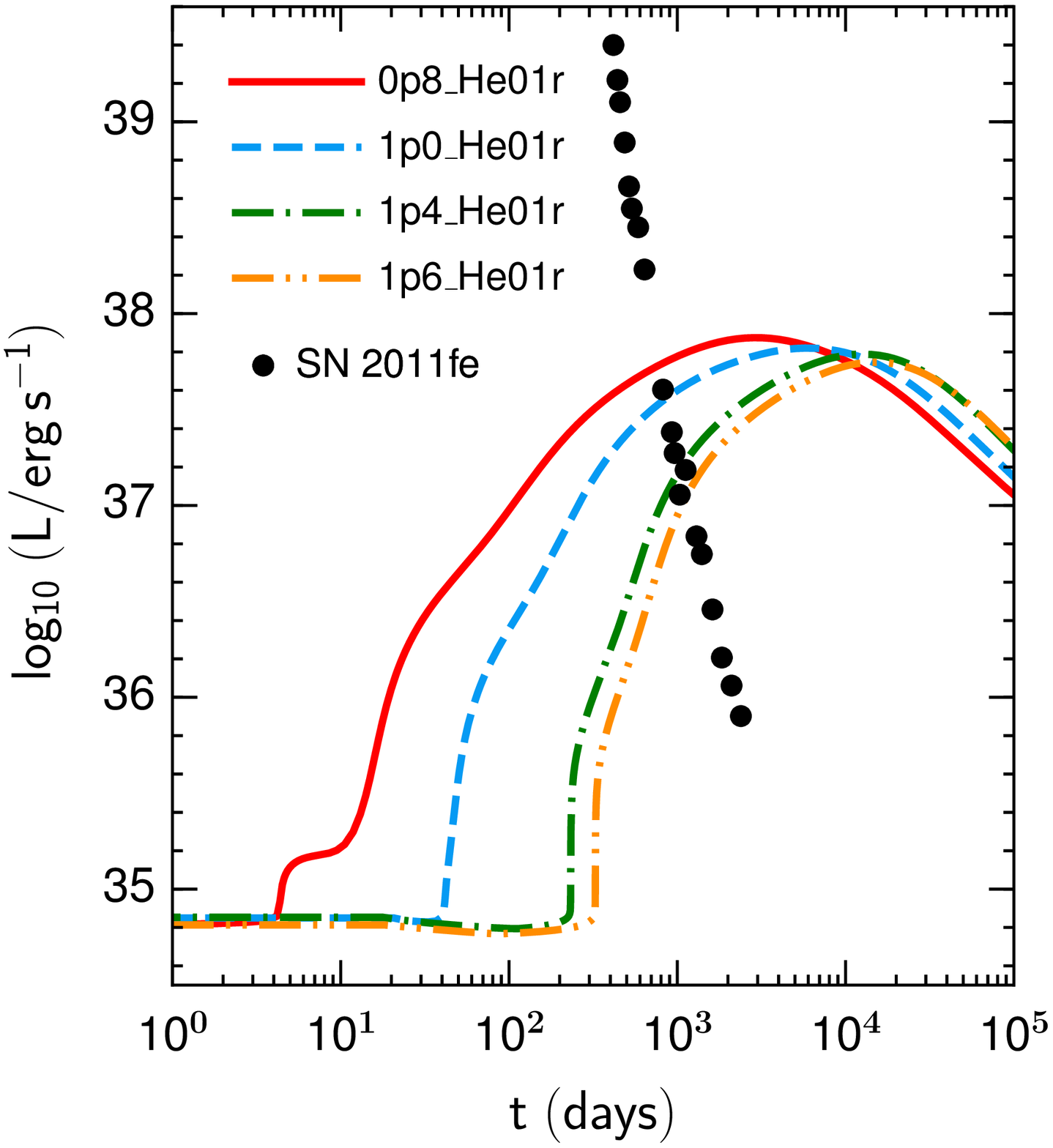}}
    \caption{\textsl{Left panel}: Comparison between predicted luminosities of our surviving He-star companion models and quasi-bolometric late-time light curve (\textsl{filled black circles}) of SN~2011fe \citep[e.g.,][]{Shappee2017,Kerzendorf2017,Dimitriadis2017,Tucker2021b}.  \textsl{Right panel}: same as \textsl{left panel}, but for the `He01r model' with different explosion energies ($0.8,1.0,1.4$ and $1.6\times10^{51}\,\mathrm{erg}$) in impact simulations (see Sect.~\ref{sec:energy}).}
\label{Fig:7}
  \end{center}
\end{figure*}

In Figure~\ref{Fig:2} we present the temporal evolution of post-impact photospheric luminosity, $L$ (panel a), the effective temperature, $T_{\mathrm{eff}}$ (panel b), the radius, $R$ (panel c), and the surface gravity, $g$ (panel d), as functions of time for the He01 and He02 models (Table~\ref{table:1}). The luminosities and photospheric radii of the two companion stars increase significantly as they expand. They reach peak luminosities of $L_{\mathrm{peak}}\sim4000$--$16000\,L_{\odot}$ at $t_{\mathrm{peak}}\sim10$--$30\,\rm{yr}$ after the explosion when the stars have radiated away the deposited energy\footnote{In this work, we define the parameters of `$L_{\mathrm{peak}}$' and  `$t_{\mathrm{peak}}$' as the post-impact peak lumonisity of the suriving companion star and the time of peak lumonisity before the star relaxes back into the thermal equilibrium.}. The stars then start to contract and enter their thermal re-equilibration phase over the Kelvin-Helmholtz timescale. After re-establishing thermal equilibrium, they evolve into a B-type or O-type subdwarf (i.e., sdB or sdO) star and follow an evolutionary track of a He star with the same mass that has not experienced ejecta-companion interaction. This suggests that the details of the interaction does not significantly affect the observational properties of the surviving He-star companion of SNe Ia after they have relaxed back into thermal equilibrium. As shown in Fig.~\ref{Fig:2}b, the effective temperature increases after the impact due to the reheating when the deposited energy diffuses out, although the star expands significantly at this moment. At some point, the expansion of the star becomes dominant, leading to a decrease of its effective temperature at about $1\,\mathrm{yr}$ after the explosion for He01 model. Subsequently, the effective temperature increases again as the star enters the Kelvin-Helmholtz contraction phase. Our findings are consistent with those of \citet[][]{Pan2013}. In addition, we note that the core He-burning in Model He01 and He-shell burning in Model He02 cause a small increase in the effective temperature and in the luminosity at around $10^{5}\,\mathrm{yr}$ after the impact (see left-hand side panel of Fig.~\ref{Fig:2}).

Figure~\ref{Fig:3} presents evolutionary tracks of our surviving He-star companions in the Hertzsprung--Russell (H--R) diagram and in the effective temperature--surface gravity ($T_{\mathrm{eff}}$--$g$) diagram. The surviving He-star companions become significantly overluminous after the impact. About a few hundred years after the explosion, the luminosity of a surviving He-star companion becomes comparable to that of a normal He-star that has not been impacted and heated. This suggests that the identification of surviving He-star companion stars of SNe Ia may be more likely to be successful for the young nearby SNRs  (which should be younger than a few $\,100\,\mathrm{yr}$).

\subsection{Influence of orbital motion and stellar spin}

To examine the dependence of numerical results of ejecta--companion interaction on the orbital motion and the spin of the He-star companion, \citet{Liu2013b} have carried out hydrodynamical impact simulation for He01 and He02 models by taking their orbital and spin velocities into account. Figs.~\ref{Fig:2} and~\ref{Fig:3} show how the orbital motion and the spin of the He-star companion affect the post-impact observational properties of a surviving He-star companion star.  As was to be expected, we find small differences between the models that include orbital motion and spin and those that do not (Table~\ref{table:1}). For instance, almost the same total amount of energy deposition into the companion star during the ejecta-companion interaction is observed in the rotating and non-rotating models. The post-impact peak luminosity $L_{\mathrm{peak}}$ and the time of peak luminosity $t_{\mathrm{peak}}$ change by a factor of about $1.1$--$1.3$ and $1.5$--$2.4$, respectively. This is in accordance with our expectations because the orbital and spin velocities (which are about $240$--$430\,\mathrm{km\,s^{-1}}$, see Table~\ref{table:1}) of a He-star companion are much lower than the typical expansion velocity of SN Ia ejecta of about $10^{4}\,\rm{km\,s^{-1}}$.

The centrifugal force is expected to make the photosphere of the rotating model expand stronger than that of the non-rotating model and therefore the luminosity would be expected to increase faster. This, however, is not what we observe: the rise is faster and luminosities reach higher values in the non-rotating models (see Fig.~\ref{Fig:2}). The reason for this counterintuitive result is that the post-impact expansion of the surviving companion star is strongly determined by the amount and depth of energy deposition. A difference between rotating and non-rotating models is seen in the depth of energy deposition into the companion star during the interaction. The depths of energy deposition (at $M_{\mathrm{r}}/M_{\ast}\sim0.98$ for the `He01' model, where $M_{\mathrm{r}}$ and $M_{\ast}$ are the enclosed mass within a sphere of radius $r$ and the total mass of the star, respectively) are shallower in non-rotating models comparing with that of rotating models (which is at $M_{\mathrm{r}}/M_{\ast}\sim0.97$ for the `He01r' model), causing a shorter local radiative diffusion timescale as seen in Fig.~\ref{Fig:2}. Therefore, the rotating models take longer to reach their maximum luminosities than the non-rotating models. For a given rotating model computed from our 3D impact simulation (i.e., for fixed other parameters such as amount and depth of the energy deposition), we have also done a test by following its post-impact evolution in \texttt{MESA} without and with including the rotation (i.e., by setting its radial angular momentum profile to be zero or not). Indeed, we find that the run including rotation has a slightly stronger expansion than the one with zero angular momentum.

\subsection{Influence of the explosion energy}
\label{sec:energy}

In \citet{Liu2013b}, the typical impact simulations were ran by adopting the so-called `W7 model' with an explosion energy of $1.23\times10^{51}\,\rm{erg}$ \citep{Nomoto1984} to represent the SN explosion. However, the exact explosion mechanism of SNe Ia remains an open question to date. Typically, different deflagration and detonation explosion models could lead to a range of kinetic energies of the ejecta spanning $0.8$--$1.6\times10^{51}\,\rm{erg}$ \citep[e.g.,][]{Roepke2007a,Roepke2007b,Seitenzahl2013}. Therefore, \citet[][see their Sec.~4.3.3]{Liu2013b} artificially adjusted the kinetic energy of the SN ejecta by scaling the velocities based on the original W7 model to investigate their effects on the ejecta-companion interaction for a given companion star model He01r (see Table~\ref{table:1}). Here, we aim to explore the dependence of the post-impact properties of a surviving He-star companion on different kinetic energies of the supernova ejecta.

Figure~\ref{Fig:4} illustrates the post-impact luminosity and effective temperature of the surviving companion star model He01r as function of time for the cases with different SN explosion energies (i.e., kinetic energies of the ejecta of $0.8$, $1.0$, $1.4$ and $1.6\times10^{51}\,\mathrm{erg}$). An increase of the ejecta energy from $0.8\times10^{51}\,\mathrm{erg}$ (model 0p8\_He01r) to $1.6\times10^{51}\,\mathrm{erg}$ (model 1p6\_He01r) leads to an increase in the amount of energy deposition into the companion star increases from $1.47\times10^{49}\,\mathrm{erg}$ to $1.88\times10^{49}\,\mathrm{erg}$ (Table~\ref{table:1}). The post-impact evolution of the surviving companion star depends strongly on the total amount and depth of energy deposition into the star during the interaction. As shown in Fig.~\ref{Fig:4}, in the model with a higher kinetic energy of the ejecta (1p6\_He01r) the surviving companion star reaches its peak luminosity later and has a lower peak luminosity during the early thermal re-equilibration phase. This is because the depth of energy deposition in 1p6\_He01r model is deeper, although the total amount of energy deposition in this model is higher. Figure~\ref{Fig:5} shows the post-impact specific internal energy of the surviving companion star as a function of mass fraction for four cases with different kinetic energies. To better show the position of energy deposition, only the outer layers of the star (i.e., $M_{\mathrm{r}}/M_{\ast}=0.8$--$1.0$) are shown in Fig.~\ref{Fig:5}. The local maximum of the internal energy seen in the envelope (i.e., the depth of energy deposition) gets deeper as the explosion energy of the ejecta increases. As a consequence, the competition between the total amount of and the depth of energy deposition leads to the results observed in Fig.~\ref{Fig:4}.

Figure~\ref{Fig:6} shows post-impact evolutionary tracks of these four models in the H--R diagram. About $10^{4}\,\mathrm{yr}$ after the explosion, our four models have re-established thermal equilibrium. Subsequently, they continue to evolve on a track quite similar to that of an unperturbed He star. This suggests that different explosion energies will not significantly affect late-time evolution of the surviving companion star after its re-equilibration. Our results are consistent with those of \citet[][]{Pan2013}.  

\subsection{Post-impact rotation}

The initial rotation of the He-star companion was set up as rigid-body rotation at the beginning of our previous 3D impact simulations \citet[][]{Liu2013b}. Under the assumption that the companion star co-rotates with its orbit due to the strong tidal interaction during the pre-explosion mass-transfer phase, the companion star models He01r and He02r have rotational surface velocities as high as  $V_{\mathrm{rot}}^{\mathrm{f}}=301\,\mathrm{km\,s^{-1}}$ and $V_{\mathrm{rot}}^{\mathrm{f}}=237\,\mathrm{km\,s^{-1}}$, respectively, at the onset of the SN explosion (see Table~\ref{table:1}). After the SN impact, the He-star companions are no longer in rigid-body rotation. We find that our He01r and He0er model lose about $13\%$ and $38\%$, respectively, of their initial angular momenta \citep[][see their Fig.~20]{Liu2013b} as  $3\%$ and $6\%$ of their masses are removed by the supernova blast wave. At the same time, the companion stars expand due to the shock heating. As a consequence, the surface rotational speeds in model He01r and He02r drop to $\sim90\,\mathrm{km\,s^{-1}}$ and $\sim50\,\mathrm{km\,s^{-1}}$, respectively, at the end of our impact SPH simulations.

Fig.~\ref{Fig:vrot} shows the post-impact evolution of the surface rotational velocities of the companions in model He01r and He02r. As expected, we find that the surface rotational velocities significantly drop as the stars expand after the impact. About $20$--$30\,\mathrm{yr}$ after the impact, when the stars reach their maximum expansion, the values decrease to about $30$--$40\,\mathrm{km\,s^{-1}}$. Subsequently, the companion stars shrink, because the energy deposited by the supernova blast wave has been radiated away. This causes the surfaces to spin up again. The stars become fast-rotating objects after thermal equilibrium is reestablished,  and we find that they rotate with surface speeds of $\sim230\,\mathrm{km\,s^{-1}}$ and $\sim190\,\mathrm{km\,s^{-1}}$ in models He01r and He02r, respectively. Our results predict that the observable rotation rate of the surviving He-star companions of SNe Ia depends on the age of the SNR. Although initially the companions were fast rotators, they rotate slowly shortly after the impact of the SN blast wave and before they start to relax back into thermal equilibrium, which again leads to an increase of the rotation speed on a timescale of $10^4\, \mathrm{yr}$.

\section{Comparison with extremely late-time observations of SN~2011\lowercase{fe}}

SN~2011fe is one of the best-observed and closest SNe Ia (its distance is estimated with $6.4\,\mathrm{Mpc}$, \citealt{Shappee2011}). This provides us with a unique opportunity to put tight constraints on the nature of its progenitor and its explosion mechanism \citep[e.g.,][]{Nugent2011,Roepke2012,Bloom2012}. Pre-explosion observations of the \textit{Hubble Space Telescope} (HST) exclude luminous red giants ($>3.5\,M_{\odot}$) and almost all He stars as companion stars in the progenitor system of SN~2011fe \citep{Li2011}. Non-detection of early-time UV and optical emission potentially caused by shock heating in the ejecta--companion collision \citep{Brown2012} and a lack of H-rich material in nebular spectra \citep[e.g.,][]{Shappee2013,Graham2015,Lundqvist2015,Tucker2021a} rule out giants and MS companions. In addition, X-ray and deep radio observations disfavor most of the popular SD progenitor models \citep[e.g.,][]{Horesh2012,Margutti2012,Chomiuk2012}.

In this section, similar to that in \citet{Shappee2013}, we use the extremely late-time (up to about 2400 days after the maximum light; e.g., \citealt[][]{Shappee2017,Dimitriadis2017,Kerzendorf2017,Tucker2021b}) photometry observations of SN~2011fe to place further constraints on the possibility of a He-star as the companion in the progenitor system of SN~2011fe. We caution that this is meant as an example and there are reasons to believe that this supernova has not resulted from a SD progenitor system \citep[e.g.,][]{Li2011,Nugent2011,Roepke2012,Brown2012,Horesh2012,Shappee2013} In the 3D impact simulation  \citep[][]{Liu2013b} underlying our current study, the `W7 model' of \citet[][]{Nomoto1984} was used for representing a normal SNe Ia explosion and we therefore select the typical normal SN~Ia SN~2011fe, which has the latest-epoch photometry observations so far as an example to make the comparison. Using an explosion model proposed for the `SNe~Iax' subclass (that is likely connected to SD progenitors) in the 3D impact simulation instead of W7, a similar comparison can be made to place constraints on the progenitor companions of sub-classes SNe Ia. Such a detailed comparison of a consistent impact model with late-time observations of a SN~Iax SN~2012Z \citep[][]{McCully2021} will be presented in our forthcoming paper (Zeng et al. in prep).

In Figure~\ref{Fig:7}, we compare the predicted luminosities of our surviving He-star companion models with the quasi-bolometric late-time light curve of SN~2011fe \citep[][]{Shappee2017,Dimitriadis2017,Kerzendorf2017,Tucker2021b}. Our shocked He-star companion models become significantly more luminous than the observed light-curve of SN~2011fe at about $1000\,\mathrm{d}$ after the explosion. This means that the very late-time light curve of SN~2011fe should be more luminous than the current observations if there is indeed a shocked surviving He-star companion. This suggests that a He-star is very unlikely the mass-donating companion to the progenitor of SN~2011fe. Our conclusions are consistent with that derived from pre-explosion observations of SN~2011fe \citep{Li2011}. 

As shown in Fig.~\ref{Fig:7}, the latest photometry observation of SN~2011fe is still slightly luminous than our `He02r' model at similar epoch, suggesting that we possibly have not yet seen a potential contribution from the shock-heated companion star with a mass of $\lesssim0.95\,M_{\odot}$ (e.g., our `He02r' model). However, about $1000\,\mathrm{d}$ after the explosion, the luminosities of our surviving companion models are still significantly increasing for a few ten years as the companions expand, and they will reach the maximum luminosity ($\gtrsim 4000\,L_{\odot}$) at about $10^{4}\,\mathrm{d}$. This means that a detectable rebrightening of the light-curve of SN~2011fe would be seen at a later epoch if there were a shocked He-star companion with a mass of $\lesssim0.95\,M_{\odot}$. Based on our comparison with current late-time observations of SN~2011fe, He-star companions with masses above $0.95\,M_{\odot}$ are ruled out. We strongly encourage future observations to test this hypothesis, which will help complete the picture of constraints on the He-star as a companion to the progenitor of SN~2011fe.

We did not consider the rotation of an accreting WD when constructing the companion star model at the moment of SN explosion in our 1D binary evolution calculation \citep{Liu2013a}. The accretion from the companion star is expected to spin up the WD, delaying its explosion (i.e., the `spin up/spin down' model; \citealt{Justham2011, DiStefano2011}). If the spin-down timescale is long enough, the He-star companion in this model could have evolved to a WD at the moment of SN explosion. In this case, the post-impact properties of the surviving companion star would be different from the results in this work because of a much weaker ejecta--companion interaction. This may lead to a dim surviving companion star that could explain the extreme late-time light-curve of SN~2011fe. However, the exact spin-down timescale of the WD in the `spin up/spin down' model is quite unknown.

\section{Summary and Conclusion}

In this work, we map the computed surviving He-star companion models from our previous 3D hydrodynamical simulations of ejecta-companion interaction \citep[][]{Liu2013a} into the 1D stellar evolution code \texttt{MESA} \citep[][]{Paxton2011, Paxton2018} to follow their subsequent re-equilibration evolution. The main goal of this work is to determine the expected observable signatures of surviving He-star companions of SNe Ia, which can be instrumental in searches for such objects in historical SN Ia remnants. Our main results and conclusions can be summarized as follows: 
\vspace{-\topsep}

\begin{itemize}\itemsep5pt 
\item[(1)] We find that the He-star companions are strongly heated and inflated during the ejecta--companion interaction. They continue to expand for a few tens of years before they start to contract. As a consequence, they become significantly overluminous over a Kelvin--Helmholtz timescale after the impact. After re-establishing thermal equilibrium, the stars continue to evolve on a track very close to that of an unperturbed He star with the same mass.

\item[(2)] Our `He01' and `He02' models absorb energies of $E_{\mathrm{in}}\sim1.68\times10^{49}\,\mathrm{erg}$ and $1.19\times10^{49}\,\mathrm{erg}$ during the ejecta--companion interaction. During the thermal re-equilibration phase, they reach a peak luminosity of $L_{\mathrm{peak}}\sim16000\,L_{\odot}$ and $5800\,L_{\odot}$ at about $t_{\mathrm{peak}}\sim15\,\mathrm{yr}$ and $30\,\mathrm{yr}$, respectively.

\item[(3)] The luminosities of our surviving He-star companions become comparable to those of a normal He-star that has not been impacted and heated a few hundred years after the explosion. This suggests that the identification of surviving He-star companions is more promising in the young nearby SNRs (younger than a few $100\,\mathrm{yr}$).

\item[(4)] We find that an inclusion of the orbital motion and the spin of the He-star companion into the impact simulation does not play an important role in determining the properties of surviving He-star companions (Figs~\ref{Fig:2}~and~\ref{Fig:3}). For example, the `He01r' model takes $t_{\mathrm{peak}}\sim22\,\mathrm{yr}$ to research its peak luminosity of $L_{\mathrm{peak}}\sim15000\,L_{\odot}$, and the two orbital parameters  ($t_{\mathrm{peak}}$ and $L_{\mathrm{peak}}$) are prolong $t_{\mathrm{peak}}$ by a factor of 1.5 and lower $L_{\mathrm{peak}}$ by 1.1 compared with model `He01' model.

\item[(5)] Artificially adjusting the kinetic energy of the SN ejecta by scaling the velocities of the original `W7 explosion model' in our impact simulations, we find that the surviving He-star companion takes a longer time to reach a lower peak luminosity during the thermal re-equilibration phase in the case with a higher kinetic energy. For a given He-star companion model `He01r', we find that the post-impact peak luminosity $L_{\mathrm{peak}}$ of the star decreases by a factor of about 1.4, and the time of peak luminosity $t_{\mathrm{peak}}$ increases by a factor of about 5, as the kinetic energy of SN Ia ejecta increases from $0.8\times10^{51}\,\mathrm{erg}$ to $1.6\times10^{51}\,\mathrm{erg}$ (Fig.~\ref{Fig:4}).

\item[(6)] The surface rotational speeds of companions in model He01r and He02r significantly decrease after the impact due to the angular momentum loss and their expansion. About $20$--$30\,\mathrm{yr}$ after the impact, they drop to around $30$--$40\,\mathrm{km\,s^{-1}}$ (i.e., $12\%$--$\%13$ of the original value) when the stars expand to the maximum size. The star then shrinks and its surface rotational speed increases again. After thermal equilibrium is reestablished, it becomes a fast-rotating object again. We suggest that the surviving He-star companions of SNe Ia could rotate slowly although they are fast-rotating stars originally, which depends on the ages of SNRs.
     
\item[(7)]  Comparing our results with the late-time light curve of the best observed SN Ia, SN~2011fe, we find that the predicted luminosities of the surviving He-star companions are generally higher significantly than the observed luminosity of SN~2011fe about $1000\,\mathrm{d}$ after the impact. This conflict seems to rule out the existence of a shocked surviving He-star companion in nearby SN~2011fe.

\item[(8)] Our results suggest that the surviving He-star companion should begin to dominate the SN Ia late-time light-curve about 1000 days after maximum light.

\item[(9)] Based on our results, there is a slight chance that current late-time light curve observations of SN~2011fe still miss a contribution from the shock-heated He-star companion that would have a mass of $\lesssim0.95\,M_{\odot}$. In this case, however, our models suggest that a rebrightening of the light-curve of SN 2011fe will be seen soon. We encourage future observations to test this hypothesis and to place further constraints on a He-star as potential companion in the progenitor system of SN~2011fe.

\end{itemize}

\begin{acknowledgments}

We thank the anonymous referee for constructive comments that helped to improve this paper. ZWL is supported by the National Key R\&D Program of China (No. 2021YFA1600401), the National Natural Science Foundation of China (NSFC, No.\ 11873016), the Chinese Academy of Sciences (CAS) and the Natural Science Foundation of Yunnan Province (No.\ 202001AW070007). The work of FR is supported by the Klaus Tschira Foundation and by the Deutsche Forschungsgemeinschaft (DFG, German Research Foundation) -- Project-ID 138713538 -- SFB 881 (``The Milky Way System'', Subproject A10). The authors gratefully acknowledge the `PHOENIX Supercomputing Platform' jointly operated by the Binary Population Synthesis Group and the Stellar Astrophysics Group at Yunnan Observatories, CAS. \\ 
\end{acknowledgments}

\vspace{5mm}

\software{\textsc{Stellar Gadget} \citep[][]{Springel2001,Springel2005,Pakmor2012a}, \texttt{MESA} \citep{Paxton2011,Paxton2013,Paxton2015,Paxton2018},
\texttt{matplotlib} \citep{Hunter2007}.  
          }

\vspace{35mm}

\bibliography{ref}{}

\begin{thebibliography}{}
\expandafter\ifx\csname natexlab\endcsname\relax\def\natexlab#1{#1}\fi
\providecommand{\url}[1]{\href{#1}{#1}}
\providecommand{\dodoi}[1]{doi:~\href{http://doi.org/#1}{\nolinkurl{#1}}}
\providecommand{\doeprint}[1]{\href{http://ascl.net/#1}{\nolinkurl{http://ascl.net/#1}}}
\providecommand{\doarXiv}[1]{\href{https://arxiv.org/abs/#1}{\nolinkurl{https://arxiv.org/abs/#1}}}

\bibitem[{{Bauer} {et~al.}(2019){Bauer}, {White}, \& {Bildsten}}]{Bauer2019}
{Bauer}, E.~B., {White}, C.~J., \& {Bildsten}, L. 2019, \apj, 887, 68,
  \dodoi{10.3847/1538-4357/ab4ea4}

\bibitem[{{Bedin} {et~al.}(2014){Bedin}, {Ruiz-Lapuente}, {Gonz{\'a}lez
  Hern{\'a}ndez}, {Canal}, {Filippenko}, \& {Mendez}}]{Bedin2014}
{Bedin}, L.~R., {Ruiz-Lapuente}, P., {Gonz{\'a}lez Hern{\'a}ndez}, J.~I.,
  {et~al.} 2014, \mnras, 439, 354, \dodoi{10.1093/mnras/stt2460}

\bibitem[{{Benz} {et~al.}(1989){Benz}, {Hills}, \& {Thielemann}}]{Benz1989}
{Benz}, W., {Hills}, J.~G., \& {Thielemann}, F.~K. 1989, \apj, 342, 986,
  \dodoi{10.1086/167656}

\bibitem[{{Bloom} {et~al.}(2012){Bloom}, {Kasen}, {Shen}, {Nugent}, {Butler},
  {Graham}, {Howell}, {Kolb}, {Holmes}, {Haswell}, {Burwitz}, {Rodriguez}, \&
  {Sullivan}}]{Bloom2012}
{Bloom}, J.~S., {Kasen}, D., {Shen}, K.~J., {et~al.} 2012, \apjl, 744, L17,
  \dodoi{10.1088/2041-8205/744/2/L17}

\bibitem[{{Boehner} {et~al.}(2017){Boehner}, {Plewa}, \&
  {Langer}}]{Boehner2017}
{Boehner}, P., {Plewa}, T., \& {Langer}, N. 2017, \mnras, 465, 2060,
  \dodoi{10.1093/mnras/stw2737}

\bibitem[{{Brown} {et~al.}(2012){Brown}, {Dawson}, {de Pasquale}, {Gronwall},
  {Holland}, {Immler}, {Kuin}, {Mazzali}, {Milne}, {Oates}, \&
  {Siegel}}]{Brown2012}
{Brown}, P.~J., {Dawson}, K.~S., {de Pasquale}, M., {et~al.} 2012, \apj, 753,
  22, \dodoi{10.1088/0004-637X/753/1/22}

\bibitem[{{Chomiuk} {et~al.}(2012){Chomiuk}, {Soderberg}, {Moe}, {Chevalier},
  {Rupen}, {Badenes}, {Margutti}, {Fransson}, {Fong}, \&
  {Dittmann}}]{Chomiuk2012}
{Chomiuk}, L., {Soderberg}, A.~M., {Moe}, M., {et~al.} 2012, \apj, 750, 164,
  \dodoi{10.1088/0004-637X/750/2/164}

\bibitem[{{Dan} {et~al.}(2011){Dan}, {Rosswog}, {Guillochon}, \&
  {Ramirez-Ruiz}}]{Dan2011}
{Dan}, M., {Rosswog}, S., {Guillochon}, J., \& {Ramirez-Ruiz}, E. 2011, \apj,
  737, 89, \dodoi{10.1088/0004-637X/737/2/89}

\bibitem[{{Di Stefano} {et~al.}(2011){Di Stefano}, {Voss}, \&
  {Claeys}}]{DiStefano2011}
{Di Stefano}, R., {Voss}, R., \& {Claeys}, J.~S.~W. 2011, \apjl, 738, L1,
  \dodoi{10.1088/2041-8205/738/1/L1}

\bibitem[{{Dimitriadis} {et~al.}(2017){Dimitriadis}, {Sullivan}, {Kerzendorf},
  {Ruiter}, {Seitenzahl}, {Taubenberger}, {Doran}, {Gal-Yam}, {Laher},
  {Maguire}, {Nugent}, {Ofek}, \& {Surace}}]{Dimitriadis2017}
{Dimitriadis}, G., {Sullivan}, M., {Kerzendorf}, W., {et~al.} 2017, \mnras,
  468, 3798, \dodoi{10.1093/mnras/stx683}

\bibitem[{{Edwards} {et~al.}(2012){Edwards}, {Pagnotta}, \&
  {Schaefer}}]{Edwards2012}
{Edwards}, Z.~I., {Pagnotta}, A., \& {Schaefer}, B.~E. 2012, \apjl, 747, L19,
  \dodoi{10.1088/2041-8205/747/2/L19}

\bibitem[{{Fink} {et~al.}(2007){Fink}, {Hillebrandt}, \&
  {R{\"o}pke}}]{Fink2007}
{Fink}, M., {Hillebrandt}, W., \& {R{\"o}pke}, F.~K. 2007, \aap, 476, 1133,
  \dodoi{10.1051/0004-6361:20078438}

\bibitem[{{Fink} {et~al.}(2014){Fink}, {Kromer}, {Seitenzahl},
  {Ciaraldi-Schoolmann}, {R{\"o}pke}, {Sim}, {Pakmor}, {Ruiter}, \&
  {Hillebrandt}}]{Fink2014}
{Fink}, M., {Kromer}, M., {Seitenzahl}, I.~R., {et~al.} 2014, \mnras, 438,
  1762, \dodoi{10.1093/mnras/stt2315}

\bibitem[{{Foley} {et~al.}(2014){Foley}, {McCully}, {Jha}, {Bildsten}, {Fong},
  {Narayan}, {Rest}, \& {Stritzinger}}]{Foley2014}
{Foley}, R.~J., {McCully}, C., {Jha}, S.~W., {et~al.} 2014, \apj, 792, 29,
  \dodoi{10.1088/0004-637X/792/1/29}

\bibitem[{{Foley} {et~al.}(2013){Foley}, {Challis}, {Chornock},
  {Ganeshalingam}, {Li}, {Marion}, {Morrell}, {Pignata}, {Stritzinger},
  {Silverman}, {Wang}, {Anderson}, {Filippenko}, {Freedman}, {Hamuy}, {Jha},
  {Kirshner}, {McCully}, {Persson}, {Phillips}, {Reichart}, \&
  {Soderberg}}]{Foley2013}
{Foley}, R.~J., {Challis}, P.~J., {Chornock}, R., {et~al.} 2013, \apj, 767, 57,
  \dodoi{10.1088/0004-637X/767/1/57}

\bibitem[{{Fuhrmann}(2005)}]{Fuhrmann2005}
{Fuhrmann}, K. 2005, \mnras, 359, L35, \dodoi{10.1111/j.1745-3933.2005.00032.x}

\bibitem[{{Gaia Collaboration} {et~al.}(2016){Gaia Collaboration}, {Prusti},
  {de Bruijne}, {Brown}, {Vallenari}, {Babusiaux}, {Bailer-Jones}, {Bastian},
  {Biermann}, {Evans}, {Eyer}, {Jansen}, {Jordi}, {Klioner}, {Lammers},
  {Lindegren}, {Luri}, {Mignard}, {Milligan}, {Panem}, {Poinsignon},
  {Pourbaix}, {Randich}, {Sarri}, {Sartoretti}, {Siddiqui}, {Soubiran},
  {Valette}, {van Leeuwen}, {Walton}, {Aerts}, {Arenou}, {Cropper}, {Drimmel},
  {H{\o}g}, {Katz}, {Lattanzi}, {O'Mullane}, {Grebel}, {Holland}, {Huc},
  {Passot}, {Bramante}, {Cacciari}, {Casta{\~n}eda}, {Chaoul}, {Cheek}, {De
  Angeli}, {Fabricius}, {Guerra}, {Hern{\'a}ndez}, {Jean-Antoine-Piccolo},
  {Masana}, {Messineo}, {Mowlavi}, {Nienartowicz}, {Ord{\'o}{\~n}ez-Blanco},
  {Panuzzo}, {Portell}, {Richards}, {Riello}, {Seabroke}, {Tanga},
  {Th{\'e}venin}, {Torra}, {Els}, {Gracia-Abril}, {Comoretto},
  {Garcia-Reinaldos}, {Lock}, {Mercier}, {Altmann}, {Andrae}, {Astraatmadja},
  {Bellas-Velidis}, {Benson}, {Berthier}, {Blomme}, {Busso}, {Carry},
  {Cellino}, {Clementini}, {Cowell}, {Creevey}, {Cuypers}, {Davidson}, {De
  Ridder}, {de Torres}, {Delchambre}, {Dell'Oro}, {Ducourant}, {Fr{\'e}mat},
  {Garc{\'\i}a-Torres}, {Gosset}, {Halbwachs}, {Hambly}, {Harrison}, {Hauser},
  {Hestroffer}, {Hodgkin}, {Huckle}, {Hutton}, {Jasniewicz}, {Jordan},
  {Kontizas}, {Korn}, {Lanzafame}, {Manteiga}, {Moitinho}, {Muinonen},
  {Osinde}, {Pancino}, {Pauwels}, {Petit}, {Recio-Blanco}, {Robin}, {Sarro},
  {Siopis}, {Smith}, {Smith}, {Sozzetti}, {Thuillot}, {van Reeven}, {Viala},
  {Abbas}, {Abreu Aramburu}, {Accart}, {Aguado}, {Allan}, {Allasia},
  {Altavilla}, {{\'A}lvarez}, {Alves}, {Anderson}, {Andrei}, {Anglada Varela},
  {Antiche}, {Antoja}, {Ant{\'o}n}, {Arcay}, {Atzei}, {Ayache}, {Bach},
  {Baker}, {Balaguer-N{\'u}{\~n}ez}, {Barache}, {Barata}, {Barbier}, {Barblan},
  {Baroni}, {Barrado y Navascu{\'e}s}, {Barros}, {Barstow}, {Becciani},
  {Bellazzini}, {Bellei}, {Bello Garc{\'\i}a}, {Belokurov}, {Bendjoya},
  {Berihuete}, {Bianchi}, {Bienaym{\'e}}, {Billebaud}, {Blagorodnova},
  {Blanco-Cuaresma}, {Boch}, {Bombrun}, {Borrachero}, {Bouquillon}, {Bourda},
  {Bouy}, {Bragaglia}, {Breddels}, {Brouillet}, {Br{\"u}semeister},
  {Bucciarelli}, {Budnik}, {Burgess}, {Burgon}, {Burlacu}, {Busonero}, {Buzzi},
  {Caffau}, {Cambras}, {Campbell}, {Cancelliere}, {Cantat-Gaudin}, {Carlucci},
  {Carrasco}, {Castellani}, {Charlot}, {Charnas}, {Charvet}, {Chassat},
  {Chiavassa}, {Clotet}, {Cocozza}, {Collins}, {Collins}, {Costigan}, {Crifo},
  {Cross}, {Crosta}, {Crowley}, {Dafonte}, {Damerdji}, {Dapergolas}, {David},
  {David}, {De Cat}, {de Felice}, {de Laverny}, {De Luise}, {De March}, {de
  Martino}, {de Souza}, {Debosscher}, {del Pozo}, {Delbo}, {Delgado},
  {Delgado}, {di Marco}, {Di Matteo}, {Diakite}, {Distefano}, {Dolding}, {Dos
  Anjos}, {Drazinos}, {Dur{\'a}n}, {Dzigan}, {Ecale}, {Edvardsson}, {Enke},
  {Erdmann}, {Escolar}, {Espina}, {Evans}, {Eynard Bontemps}, {Fabre},
  {Fabrizio}, {Faigler}, {Falc{\~a}o}, {Farr{\`a}s Casas}, {Faye}, {Federici},
  {Fedorets}, {Fern{\'a}ndez-Hern{\'a}ndez}, {Fernique}, {Fienga}, {Figueras},
  {Filippi}, {Findeisen}, {Fonti}, {Fouesneau}, {Fraile}, {Fraser}, {Fuchs},
  {Furnell}, {Gai}, {Galleti}, {Galluccio}, {Garabato}, {Garc{\'\i}a-Sedano},
  {Gar{\'e}}, {Garofalo}, {Garralda}, {Gavras}, {Gerssen}, {Geyer}, {Gilmore},
  {Girona}, {Giuffrida}, {Gomes}, {Gonz{\'a}lez-Marcos},
  {Gonz{\'a}lez-N{\'u}{\~n}ez}, {Gonz{\'a}lez-Vidal}, {Granvik}, {Guerrier},
  {Guillout}, {Guiraud}, {G{\'u}rpide}, {Guti{\'e}rrez-S{\'a}nchez}, {Guy},
  {Haigron}, {Hatzidimitriou}, {Haywood}, {Heiter}, {Helmi}, {Hobbs},
  {Hofmann}, {Holl}, {Holland}, {Hunt}, {Hypki}, {Icardi}, {Irwin}, {Jevardat
  de Fombelle}, {Jofr{\'e}}, {Jonker}, {Jorissen}, {Julbe}, {Karampelas},
  {Kochoska}, {Kohley}, {Kolenberg}, {Kontizas}, {Koposov}, {Kordopatis},
  {Koubsky}, {Kowalczyk}, {Krone-Martins}, {Kudryashova}, {Kull}, {Bachchan},
  {Lacoste-Seris}, {Lanza}, {Lavigne}, {Le Poncin-Lafitte}, {Lebreton},
  {Lebzelter}, {Leccia}, {Leclerc}, {Lecoeur-Taibi}, {Lemaitre}, {Lenhardt},
  {Leroux}, {Liao}, {Licata}, {Lindstr{\o}m}, {Lister}, {Livanou}, {Lobel},
  {L{\"o}ffler}, {L{\'o}pez}, {Lopez-Lozano}, {Lorenz}, {Loureiro},
  {MacDonald}, {Magalh{\~a}es Fernandes}, {Managau}, {Mann}, {Mantelet},
  {Marchal}, {Marchant}, {Marconi}, {Marie}, {Marinoni}, {Marrese},
  {Marschalk{\'o}}, {Marshall}, {Mart{\'\i}n-Fleitas}, {Martino}, {Mary},
  {Matijevi{\v{c}}}, {Mazeh}, {McMillan}, {Messina}, {Mestre}, {Michalik},
  {Millar}, {Miranda}, {Molina}, {Molinaro}, {Molinaro}, {Moln{\'a}r},
  {Moniez}, {Montegriffo}, {Monteiro}, {Mor}, {Mora}, {Morbidelli}, {Morel},
  {Morgenthaler}, {Morley}, {Morris}, {Mulone}, {Muraveva}, {Musella},
  {Narbonne}, {Nelemans}, {Nicastro}, {Noval}, {Ord{\'e}novic},
  {Ordieres-Mer{\'e}}, {Osborne}, {Pagani}, {Pagano}, {Pailler}, {Palacin},
  {Palaversa}, {Parsons}, {Paulsen}, {Pecoraro}, {Pedrosa}, {Pentik{\"a}inen},
  {Pereira}, {Pichon}, {Piersimoni}, {Pineau}, {Plachy}, {Plum}, {Poujoulet},
  {Pr{\v{s}}a}, {Pulone}, {Ragaini}, {Rago}, {Rambaux}, {Ramos-Lerate},
  {Ranalli}, {Rauw}, {Read}, {Regibo}, {Renk}, {Reyl{\'e}}, {Ribeiro},
  {Rimoldini}, {Ripepi}, {Riva}, {Rixon}, {Roelens}, {Romero-G{\'o}mez},
  {Rowell}, {Royer}, {Rudolph}, {Ruiz-Dern}, {Sadowski}, {Sagrist{\`a}
  Sell{\'e}s}, {Sahlmann}, {Salgado}, {Salguero}, {Sarasso}, {Savietto},
  {Schnorhk}, {Schultheis}, {Sciacca}, {Segol}, {Segovia}, {Segransan},
  {Serpell}, {Shih}, {Smareglia}, {Smart}, {Smith}, {Solano}, {Solitro},
  {Sordo}, {Soria Nieto}, {Souchay}, {Spagna}, {Spoto}, {Stampa}, {Steele},
  {Steidelm{\"u}ller}, {Stephenson}, {Stoev}, {Suess}, {S{\"u}veges}, {Surdej},
  {Szabados}, {Szegedi-Elek}, {Tapiador}, {Taris}, {Tauran}, {Taylor},
  {Teixeira}, {Terrett}, {Tingley}, {Trager}, {Turon}, {Ulla}, {Utrilla},
  {Valentini}, {van Elteren}, {Van Hemelryck}, {van Leeuwen}, {Varadi},
  {Vecchiato}, {Veljanoski}, {Via}, {Vicente}, {Vogt}, {Voss}, {Votruba},
  {Voutsinas}, {Walmsley}, {Weiler}, {Weingrill}, {Werner}, {Wevers},
  {Whitehead}, {Wyrzykowski}, {Yoldas}, {{\v{Z}}erjal}, {Zucker}, {Zurbach},
  {Zwitter}, {Alecu}, {Allen}, {Allende Prieto}, {Amorim},
  {Anglada-Escud{\'e}}, {Arsenijevic}, {Azaz}, {Balm}, {Beck}, {Bernstein},
  {Bigot}, {Bijaoui}, {Blasco}, {Bonfigli}, {Bono}, {Boudreault}, {Bressan},
  {Brown}, {Brunet}, {Bunclark}, {Buonanno}, {Butkevich}, {Carret}, {Carrion},
  {Chemin}, {Ch{\'e}reau}, {Corcione}, {Darmigny}, {de Boer}, {de Teodoro}, {de
  Zeeuw}, {Delle Luche}, {Domingues}, {Dubath}, {Fodor}, {Fr{\'e}zouls},
  {Fries}, {Fustes}, {Fyfe}, {Gallardo}, {Gallegos}, {Gardiol}, {Gebran},
  {Gomboc}, {G{\'o}mez}, {Grux}, {Gueguen}, {Heyrovsky}, {Hoar}, {Iannicola},
  {Isasi Parache}, {Janotto}, {Joliet}, {Jonckheere}, {Keil}, {Kim},
  {Klagyivik}, {Klar}, {Knude}, {Kochukhov}, {Kolka}, {Kos}, {Kutka}, {Lainey},
  {LeBouquin}, {Liu}, {Loreggia}, {Makarov}, {Marseille}, {Martayan},
  {Martinez-Rubi}, {Massart}, {Meynadier}, {Mignot}, {Munari}, {Nguyen},
  {Nordlander}, {Ocvirk}, {O'Flaherty}, {Olias Sanz}, {Ortiz}, {Osorio},
  {Oszkiewicz}, {Ouzounis}, {Palmer}, {Park}, {Pasquato}, {Peltzer}, {Peralta},
  {P{\'e}turaud}, {Pieniluoma}, {Pigozzi}, {Poels}, {Prat}, {Prod'homme},
  {Raison}, {Rebordao}, {Risquez}, {Rocca-Volmerange}, {Rosen}, {Ruiz-Fuertes},
  {Russo}, {Sembay}, {Serraller Vizcaino}, {Short}, {Siebert}, {Silva},
  {Sinachopoulos}, {Slezak}, {Soffel}, {Sosnowska}, {Strai{\v{z}}ys}, {ter
  Linden}, {Terrell}, {Theil}, {Tiede}, {Troisi}, {Tsalmantza}, {Tur},
  {Vaccari}, {Vachier}, {Valles}, {Van Hamme}, {Veltz}, {Virtanen}, {Wallut},
  {Wichmann}, {Wilkinson}, {Ziaeepour}, \& {Zschocke}}]{Gaia2016}
{Gaia Collaboration}, {Prusti}, T., {de Bruijne}, J.~H.~J., {et~al.} 2016,
  \aap, 595, A1, \dodoi{10.1051/0004-6361/201629272}

\bibitem[{{Gaia Collaboration} {et~al.}(2018){Gaia Collaboration}, {Brown},
  {Vallenari}, {Prusti}, {de Bruijne}, {Babusiaux}, {Bailer-Jones}, {Biermann},
  {Evans}, {Eyer}, {Jansen}, {Jordi}, {Klioner}, {Lammers}, {Lindegren},
  {Luri}, {Mignard}, {Panem}, {Pourbaix}, {Randich}, {Sartoretti}, {Siddiqui},
  {Soubiran}, {van Leeuwen}, {Walton}, {Arenou}, {Bastian}, {Cropper},
  {Drimmel}, {Katz}, {Lattanzi}, {Bakker}, {Cacciari}, {Casta{\~n}eda},
  {Chaoul}, {Cheek}, {De Angeli}, {Fabricius}, {Guerra}, {Holl}, {Masana},
  {Messineo}, {Mowlavi}, {Nienartowicz}, {Panuzzo}, {Portell}, {Riello},
  {Seabroke}, {Tanga}, {Th{\'e}venin}, {Gracia-Abril}, {Comoretto},
  {Garcia-Reinaldos}, {Teyssier}, {Altmann}, {Andrae}, {Audard},
  {Bellas-Velidis}, {Benson}, {Berthier}, {Blomme}, {Burgess}, {Busso},
  {Carry}, {Cellino}, {Clementini}, {Clotet}, {Creevey}, {Davidson}, {De
  Ridder}, {Delchambre}, {Dell'Oro}, {Ducourant},
  {Fern{\'a}ndez-Hern{\'a}ndez}, {Fouesneau}, {Fr{\'e}mat}, {Galluccio},
  {Garc{\'\i}a-Torres}, {Gonz{\'a}lez-N{\'u}{\~n}ez}, {Gonz{\'a}lez-Vidal},
  {Gosset}, {Guy}, {Halbwachs}, {Hambly}, {Harrison}, {Hern{\'a}ndez},
  {Hestroffer}, {Hodgkin}, {Hutton}, {Jasniewicz}, {Jean-Antoine-Piccolo},
  {Jordan}, {Korn}, {Krone-Martins}, {Lanzafame}, {Lebzelter}, {L{\"o}ffler},
  {Manteiga}, {Marrese}, {Mart{\'\i}n-Fleitas}, {Moitinho}, {Mora}, {Muinonen},
  {Osinde}, {Pancino}, {Pauwels}, {Petit}, {Recio-Blanco}, {Richards},
  {Rimoldini}, {Robin}, {Sarro}, {Siopis}, {Smith}, {Sozzetti}, {S{\"u}veges},
  {Torra}, {van Reeven}, {Abbas}, {Abreu Aramburu}, {Accart}, {Aerts},
  {Altavilla}, {{\'A}lvarez}, {Alvarez}, {Alves}, {Anderson}, {Andrei},
  {Anglada Varela}, {Antiche}, {Antoja}, {Arcay}, {Astraatmadja}, {Bach},
  {Baker}, {Balaguer-N{\'u}{\~n}ez}, {Balm}, {Barache}, {Barata}, {Barbato},
  {Barblan}, {Barklem}, {Barrado}, {Barros}, {Barstow}, {Bartholom{\'e}
  Mu{\~n}oz}, {Bassilana}, {Becciani}, {Bellazzini}, {Berihuete}, {Bertone},
  {Bianchi}, {Bienaym{\'e}}, {Blanco-Cuaresma}, {Boch}, {Boeche}, {Bombrun},
  {Borrachero}, {Bossini}, {Bouquillon}, {Bourda}, {Bragaglia}, {Bramante},
  {Breddels}, {Bressan}, {Brouillet}, {Br{\"u}semeister}, {Brugaletta},
  {Bucciarelli}, {Burlacu}, {Busonero}, {Butkevich}, {Buzzi}, {Caffau},
  {Cancelliere}, {Cannizzaro}, {Cantat-Gaudin}, {Carballo}, {Carlucci},
  {Carrasco}, {Casamiquela}, {Castellani}, {Castro-Ginard}, {Charlot},
  {Chemin}, {Chiavassa}, {Cocozza}, {Costigan}, {Cowell}, {Crifo}, {Crosta},
  {Crowley}, {Cuypers}, {Dafonte}, {Damerdji}, {Dapergolas}, {David}, {David},
  {de Laverny}, {De Luise}, {De March}, {de Martino}, {de Souza}, {de Torres},
  {Debosscher}, {del Pozo}, {Delbo}, {Delgado}, {Delgado}, {Di Matteo},
  {Diakite}, {Diener}, {Distefano}, {Dolding}, {Drazinos}, {Dur{\'a}n},
  {Edvardsson}, {Enke}, {Eriksson}, {Esquej}, {Eynard Bontemps}, {Fabre},
  {Fabrizio}, {Faigler}, {Falc{\~a}o}, {Farr{\`a}s Casas}, {Federici},
  {Fedorets}, {Fernique}, {Figueras}, {Filippi}, {Findeisen}, {Fonti},
  {Fraile}, {Fraser}, {Fr{\'e}zouls}, {Gai}, {Galleti}, {Garabato},
  {Garc{\'\i}a-Sedano}, {Garofalo}, {Garralda}, {Gavel}, {Gavras}, {Gerssen},
  {Geyer}, {Giacobbe}, {Gilmore}, {Girona}, {Giuffrida}, {Glass}, {Gomes},
  {Granvik}, {Gueguen}, {Guerrier}, {Guiraud}, {Guti{\'e}rrez-S{\'a}nchez},
  {Haigron}, {Hatzidimitriou}, {Hauser}, {Haywood}, {Heiter}, {Helmi}, {Heu},
  {Hilger}, {Hobbs}, {Hofmann}, {Holland}, {Huckle}, {Hypki}, {Icardi},
  {Jan{\ss}en}, {Jevardat de Fombelle}, {Jonker}, {Juh{\'a}sz}, {Julbe},
  {Karampelas}, {Kewley}, {Klar}, {Kochoska}, {Kohley}, {Kolenberg},
  {Kontizas}, {Kontizas}, {Koposov}, {Kordopatis}, {Kostrzewa-Rutkowska},
  {Koubsky}, {Lambert}, {Lanza}, {Lasne}, {Lavigne}, {Le Fustec}, {Le
  Poncin-Lafitte}, {Lebreton}, {Leccia}, {Leclerc}, {Lecoeur-Taibi},
  {Lenhardt}, {Leroux}, {Liao}, {Licata}, {Lindstr{\o}m}, {Lister}, {Livanou},
  {Lobel}, {L{\'o}pez}, {Managau}, {Mann}, {Mantelet}, {Marchal}, {Marchant},
  {Marconi}, {Marinoni}, {Marschalk{\'o}}, {Marshall}, {Martino}, {Marton},
  {Mary}, {Massari}, {Matijevi{\v{c}}}, {Mazeh}, {McMillan}, {Messina},
  {Michalik}, {Millar}, {Molina}, {Molinaro}, {Moln{\'a}r}, {Montegriffo},
  {Mor}, {Morbidelli}, {Morel}, {Morris}, {Mulone}, {Muraveva}, {Musella},
  {Nelemans}, {Nicastro}, {Noval}, {O'Mullane}, {Ord{\'e}novic},
  {Ord{\'o}{\~n}ez-Blanco}, {Osborne}, {Pagani}, {Pagano}, {Pailler},
  {Palacin}, {Palaversa}, {Panahi}, {Pawlak}, {Piersimoni}, {Pineau}, {Plachy},
  {Plum}, {Poggio}, {Poujoulet}, {Pr{\v{s}}a}, {Pulone}, {Racero}, {Ragaini},
  {Rambaux}, {Ramos-Lerate}, {Regibo}, {Reyl{\'e}}, {Riclet}, {Ripepi}, {Riva},
  {Rivard}, {Rixon}, {Roegiers}, {Roelens}, {Romero-G{\'o}mez}, {Rowell},
  {Royer}, {Ruiz-Dern}, {Sadowski}, {Sagrist{\`a} Sell{\'e}s}, {Sahlmann},
  {Salgado}, {Salguero}, {Sanna}, {Santana-Ros}, {Sarasso}, {Savietto},
  {Schultheis}, {Sciacca}, {Segol}, {Segovia}, {S{\'e}gransan}, {Shih},
  {Siltala}, {Silva}, {Smart}, {Smith}, {Solano}, {Solitro}, {Sordo}, {Soria
  Nieto}, {Souchay}, {Spagna}, {Spoto}, {Stampa}, {Steele},
  {Steidelm{\"u}ller}, {Stephenson}, {Stoev}, {Suess}, {Surdej}, {Szabados},
  {Szegedi-Elek}, {Tapiador}, {Taris}, {Tauran}, {Taylor}, {Teixeira},
  {Terrett}, {Teyssandier}, {Thuillot}, {Titarenko}, {Torra Clotet}, {Turon},
  {Ulla}, {Utrilla}, {Uzzi}, {Vaillant}, {Valentini}, {Valette}, {van Elteren},
  {Van Hemelryck}, {van Leeuwen}, {Vaschetto}, {Vecchiato}, {Veljanoski},
  {Viala}, {Vicente}, {Vogt}, {von Essen}, {Voss}, {Votruba}, {Voutsinas},
  {Walmsley}, {Weiler}, {Wertz}, {Wevers}, {Wyrzykowski}, {Yoldas},
  {{\v{Z}}erjal}, {Ziaeepour}, {Zorec}, {Zschocke}, {Zucker}, {Zurbach}, \&
  {Zwitter}}]{Gaia2018}
{Gaia Collaboration}, {Brown}, A.~G.~A., {Vallenari}, A., {et~al.} 2018, \aap,
  616, A1, \dodoi{10.1051/0004-6361/201833051}

\bibitem[{{Gonz{\'a}lez Hern{\'a}ndez} {et~al.}(2009){Gonz{\'a}lez
  Hern{\'a}ndez}, {Ruiz-Lapuente}, {Filippenko}, {Foley}, {Gal-Yam}, \&
  {Simon}}]{Gonzalez2009}
{Gonz{\'a}lez Hern{\'a}ndez}, J.~I., {Ruiz-Lapuente}, P., {Filippenko}, A.~V.,
  {et~al.} 2009, \apj, 691, 1, \dodoi{10.1088/0004-637X/691/1/1}

\bibitem[{{Gonz{\'a}lez Hern{\'a}ndez} {et~al.}(2012){Gonz{\'a}lez
  Hern{\'a}ndez}, {Ruiz-Lapuente}, {Tabernero}, {Montes}, {Canal},
  {M{\'e}ndez}, \& {Bedin}}]{Gonzalez2012}
{Gonz{\'a}lez Hern{\'a}ndez}, J.~I., {Ruiz-Lapuente}, P., {Tabernero}, H.~M.,
  {et~al.} 2012, \nat, 489, 533, \dodoi{10.1038/nature11447}

\bibitem[{{Graham} {et~al.}(2015){Graham}, {Nugent}, {Sullivan}, {Filippenko},
  {Cenko}, {Silverman}, {Clubb}, \& {Zheng}}]{Graham2015}
{Graham}, M.~L., {Nugent}, P.~E., {Sullivan}, M., {et~al.} 2015, \mnras, 454,
  1948, \dodoi{10.1093/mnras/stv1888}

\bibitem[{{Gronow} {et~al.}(2020){Gronow}, {Collins}, {Ohlmann}, {Pakmor},
  {Kromer}, {Seitenzahl}, {Sim}, \& {R{\"o}pke}}]{Gronow2020}
{Gronow}, S., {Collins}, C., {Ohlmann}, S.~T., {et~al.} 2020, \aap, 635, A169,
  \dodoi{10.1051/0004-6361/201936494}

\bibitem[{{Gronow} {et~al.}(2021){Gronow}, {Collins}, {Sim}, \&
  {Roepke}}]{Gronow2021}
{Gronow}, S., {Collins}, C.~E., {Sim}, S.~A., \& {Roepke}, F.~K. 2021, arXiv
  e-prints, arXiv:2102.06719.
\newblock \doarXiv{2102.06719}

\bibitem[{{Hachisu} {et~al.}(1996){Hachisu}, {Kato}, \& {Nomoto}}]{Hachisu1996}
{Hachisu}, I., {Kato}, M., \& {Nomoto}, K. 1996, \apjl, 470, L97,
  \dodoi{10.1086/310303}

\bibitem[{{Han}(2008)}]{Han2008}
{Han}, Z. 2008, \apjl, 677, L109, \dodoi{10.1086/588191}

\bibitem[{{Han} \& {Podsiadlowski}(2004)}]{Han2004}
{Han}, Z., \& {Podsiadlowski}, P. 2004, \mnras, 350, 1301,
  \dodoi{10.1111/j.1365-2966.2004.07713.x}

\bibitem[{{Heger} \& {Woosley}(2010)}]{Heger2010}
{Heger}, A., \& {Woosley}, S.~E. 2010, \apj, 724, 341,
  \dodoi{10.1088/0004-637X/724/1/341}

\bibitem[{{Hillebrandt} {et~al.}(2013){Hillebrandt}, {Kromer}, {R{\"o}pke}, \&
  {Ruiter}}]{Hillebrandt2013}
{Hillebrandt}, W., {Kromer}, M., {R{\"o}pke}, F.~K., \& {Ruiter}, A.~J. 2013,
  Frontiers of Physics, 8, 116, \dodoi{10.1007/s11467-013-0303-2}

\bibitem[{{Horesh} {et~al.}(2012){Horesh}, {Kulkarni}, {Fox}, {Carpenter},
  {Kasliwal}, {Ofek}, {Quimby}, {Gal-Yam}, {Cenko}, {de Bruyn}, {Kamble},
  {Wijers}, {van der Horst}, {Kouveliotou}, {Podsiadlowski}, {Sullivan},
  {Maguire}, {Howell}, {Nugent}, {Gehrels}, {Law}, {Poznanski}, \&
  {Shara}}]{Horesh2012}
{Horesh}, A., {Kulkarni}, S.~R., {Fox}, D.~B., {et~al.} 2012, \apj, 746, 21,
  \dodoi{10.1088/0004-637X/746/1/21}

\bibitem[{{Hoyle} \& {Fowler}(1960)}]{Hoyle1960}
{Hoyle}, F., \& {Fowler}, W.~A. 1960, \apj, 132, 565, \dodoi{10.1086/146963}

\bibitem[{Hunter(2007)}]{Hunter2007}
Hunter, J.~D. 2007, Computing in Science \& Engineering, 9, 90,
  \dodoi{10.1109/MCSE.2007.55}

\bibitem[{{Iben} \& {Tutukov}(1984)}]{Iben1984}
{Iben}, Jr., I., \& {Tutukov}, A.~V. 1984, \apj, 284, 719,
  \dodoi{10.1086/162455}

\bibitem[{{Ihara} {et~al.}(2007){Ihara}, {Ozaki}, {Doi}, {Shigeyama},
  {Kashikawa}, {Komiyama}, \& {Hattori}}]{Ihara2007}
{Ihara}, Y., {Ozaki}, J., {Doi}, M., {et~al.} 2007, \pasj, 59, 811,
  \dodoi{10.1093/pasj/59.4.811}

\bibitem[{{Ilkov} \& {Soker}(2012)}]{Ilkov2012}
{Ilkov}, M., \& {Soker}, N. 2012, \mnras, 419, 1695,
  \dodoi{10.1111/j.1365-2966.2011.19833.x}

\bibitem[{{Jordan} {et~al.}(2012){Jordan}, {Perets}, {Fisher}, \& {van
  Rossum}}]{Jordan2012}
{Jordan}, IV, G.~C., {Perets}, H.~B., {Fisher}, R.~T., \& {van Rossum}, D.~R.
  2012, \apjl, 761, L23, \dodoi{10.1088/2041-8205/761/2/L23}

\bibitem[{{Justham}(2011)}]{Justham2011}
{Justham}, S. 2011, \apjl, 730, L34+, \dodoi{10.1088/2041-8205/730/2/L34}

\bibitem[{{Kelly} {et~al.}(2014){Kelly}, {Fox}, {Filippenko}, {Cenko}, {Prato},
  {Schaefer}, {Shen}, {Zheng}, {Graham}, \& {Tucker}}]{Kelly2014}
{Kelly}, P.~L., {Fox}, O.~D., {Filippenko}, A.~V., {et~al.} 2014, \apj, 790, 3,
  \dodoi{10.1088/0004-637X/790/1/3}

\bibitem[{{Kerzendorf} {et~al.}(2014){Kerzendorf}, {Childress},
  {Scharw{\"a}chter}, {Do}, \& {Schmidt}}]{Kerzendorf2014}
{Kerzendorf}, W.~E., {Childress}, M., {Scharw{\"a}chter}, J., {Do}, T., \&
  {Schmidt}, B.~P. 2014, \apj, 782, 27, \dodoi{10.1088/0004-637X/782/1/27}

\bibitem[{{Kerzendorf} {et~al.}(2018{\natexlab{a}}){Kerzendorf}, {Long},
  {Winkler}, \& {Do}}]{Kerzendorf2018a}
{Kerzendorf}, W.~E., {Long}, K.~S., {Winkler}, P.~F., \& {Do}, T.
  2018{\natexlab{a}}, \mnras, 479, 5696, \dodoi{10.1093/mnras/sty1863}

\bibitem[{{Kerzendorf} {et~al.}(2009){Kerzendorf}, {Schmidt}, {Asplund},
  {Nomoto}, {Podsiadlowski}, {Frebel}, {Fesen}, \& {Yong}}]{Kerzendorf2009}
{Kerzendorf}, W.~E., {Schmidt}, B.~P., {Asplund}, M., {et~al.} 2009, \apj, 701,
  1665, \dodoi{10.1088/0004-637X/701/2/1665}

\bibitem[{{Kerzendorf} {et~al.}(2012){Kerzendorf}, {Schmidt}, {Laird},
  {Podsiadlowski}, \& {Bessell}}]{Kerzendorf2012}
{Kerzendorf}, W.~E., {Schmidt}, B.~P., {Laird}, J.~B., {Podsiadlowski}, P., \&
  {Bessell}, M.~S. 2012, \apj, 759, 7, \dodoi{10.1088/0004-637X/759/1/7}

\bibitem[{{Kerzendorf} {et~al.}(2018{\natexlab{b}}){Kerzendorf}, {Strampelli},
  {Shen}, {Schwab}, {Pakmor}, {Do}, {Buchner}, \& {Rest}}]{Kerzendorf2018b}
{Kerzendorf}, W.~E., {Strampelli}, G., {Shen}, K.~J., {et~al.}
  2018{\natexlab{b}}, \mnras, 479, 192, \dodoi{10.1093/mnras/sty1357}

\bibitem[{{Kerzendorf} {et~al.}(2013){Kerzendorf}, {Yong}, {Schmidt}, {Simon},
  {Jeffery}, {Anderson}, {Podsiadlowski}, {Gal-Yam}, {Silverman}, {Filippenko},
  {Nomoto}, {Murphy}, {Bessell}, {Venn}, \& {Foley}}]{Kerzendorf2013}
{Kerzendorf}, W.~E., {Yong}, D., {Schmidt}, B.~P., {et~al.} 2013, \apj, 774,
  99, \dodoi{10.1088/0004-637X/774/2/99}

\bibitem[{{Kerzendorf} {et~al.}(2017){Kerzendorf}, {McCully}, {Taubenberger},
  {Jerkstrand}, {Seitenzahl}, {Ruiter}, {Spyromilio}, {Long}, \&
  {Fransson}}]{Kerzendorf2017}
{Kerzendorf}, W.~E., {McCully}, C., {Taubenberger}, S., {et~al.} 2017, \mnras,
  472, 2534, \dodoi{10.1093/mnras/stx1923}

\bibitem[{{Kromer} {et~al.}(2013){Kromer}, {Fink}, {Stanishev}, {Taubenberger},
  {Ciaraldi-Schoolman}, {Pakmor}, {R{\"o}pke}, {Ruiter}, {Seitenzahl}, {Sim},
  {Blanc}, {Elias-Rosa}, \& {Hillebrandt}}]{Kromer2013}
{Kromer}, M., {Fink}, M., {Stanishev}, V., {et~al.} 2013, \mnras, 429, 2287,
  \dodoi{10.1093/mnras/sts498}

\bibitem[{{Kushnir} {et~al.}(2013){Kushnir}, {Katz}, {Dong}, {Livne}, \&
  {Fern{\'a}ndez}}]{Kushnir2013}
{Kushnir}, D., {Katz}, B., {Dong}, S., {Livne}, E., \& {Fern{\'a}ndez}, R.
  2013, \apjl, 778, L37, \dodoi{10.1088/2041-8205/778/2/L37}

\bibitem[{{Lach} {et~al.}(2021{\natexlab{a}}){Lach}, {Callan}, {Bubeck},
  {Roepke}, {Sim}, {Schrauth}, {Ohlmann}, \& {Kromer}}]{Lach2021}
{Lach}, F., {Callan}, F.~P., {Bubeck}, D., {et~al.} 2021{\natexlab{a}}, arXiv
  e-prints, arXiv:2109.02926.
\newblock \doarXiv{2109.02926}

\bibitem[{{Lach} {et~al.}(2021{\natexlab{b}}){Lach}, {Callan}, {Sim}, \&
  {Roepke}}]{Lach2021b}
{Lach}, F., {Callan}, F.~P., {Sim}, S.~A., \& {Roepke}, F.~K.
  2021{\natexlab{b}}, arXiv e-prints, arXiv:2111.14394.
\newblock \doarXiv{2111.14394}

\bibitem[{{Li} {et~al.}(2017){Li}, {Chu}, {Gruendl}, {Weisz}, {Pan}, {Points},
  {Ricker}, {Smith}, \& {Walter}}]{Li2017}
{Li}, C.-J., {Chu}, Y.-H., {Gruendl}, R.~A., {et~al.} 2017, \apj, 836, 85,
  \dodoi{10.3847/1538-4357/836/1/85}

\bibitem[{{Li} {et~al.}(2019){Li}, {Kerzendorf}, {Chu}, {Chen}, {Do},
  {Gruendl}, {Holmes}, {Ishioka}, {Leibundgut}, {Pan}, {Ricker}, \&
  {Weisz}}]{Li2019}
{Li}, C.-J., {Kerzendorf}, W.~E., {Chu}, Y.-H., {et~al.} 2019, \apj, 886, 99,
  \dodoi{10.3847/1538-4357/ab4a03}

\bibitem[{{Li} {et~al.}(2011){Li}, {Bloom}, {Podsiadlowski}, {Miller}, {Cenko},
  {Jha}, {Sullivan}, {Howell}, {Nugent}, {Butler}, {Ofek}, {Kasliwal},
  {Richards}, {Stockton}, {Shih}, {Bildsten}, {Shara}, {Bibby}, {Filippenko},
  {Ganeshalingam}, {Silverman}, {Kulkarni}, {Law}, {Poznanski}, {Quimby},
  {McCully}, {Patel}, {Maguire}, \& {Shen}}]{Li2011}
{Li}, W., {Bloom}, J.~S., {Podsiadlowski}, P., {et~al.} 2011, \nat, 480, 348,
  \dodoi{10.1038/nature10646}

\bibitem[{{Litke} {et~al.}(2017){Litke}, {Chu}, {Holmes}, {Santucci},
  {Blindauer}, {Gruendl}, {Li}, {Pan}, {Ricker}, \& {Weisz}}]{Litke2017}
{Litke}, K.~C., {Chu}, Y.-H., {Holmes}, A., {et~al.} 2017, \apj, 837, 111,
  \dodoi{10.3847/1538-4357/aa5d57}

\bibitem[{{Liu} \& {Stancliffe}(2020)}]{Liu2020aa}
{Liu}, Z., \& {Stancliffe}, R.~J. 2020, \aap, 641, A20,
  \dodoi{10.1051/0004-6361/202038443}

\bibitem[{{Liu} {et~al.}(2013{\natexlab{a}}){Liu}, {Kromer}, {Fink}, {Pakmor},
  {R{\"o}pke}, {Chen}, {Wang}, \& {Han}}]{Liu2013c}
{Liu}, Z.-W., {Kromer}, M., {Fink}, M., {et~al.} 2013{\natexlab{a}}, \apj, 778,
  121, \dodoi{10.1088/0004-637X/778/2/121}

\bibitem[{{Liu} {et~al.}(2013{\natexlab{b}}){Liu}, {Pakmor}, {R{\"o}pke},
  {Edelmann}, {Hillebrandt}, {Kerzendorf}, {Wang}, \& {Han}}]{Liu2013a}
{Liu}, Z.-W., {Pakmor}, R., {R{\"o}pke}, F.~K., {et~al.} 2013{\natexlab{b}},
  \aap, 554, A109, \dodoi{10.1051/0004-6361/201220903}

\bibitem[{{Liu} {et~al.}(2012){Liu}, {Pakmor}, {R{\"o}pke}, {Edelmann}, {Wang},
  {Kromer}, {Hillebrandt}, \& {Han}}]{Liu2012}
{Liu}, Z.~W., {Pakmor}, R., {R{\"o}pke}, F.~K., {et~al.} 2012, \aap, 548, A2,
  \dodoi{10.1051/0004-6361/201219357}

\bibitem[{{Liu} {et~al.}(2021){Liu}, {R{\"o}pke}, {Zeng}, \&
  {Heger}}]{Liu2021b}
{Liu}, Z.-W., {R{\"o}pke}, F.~K., {Zeng}, Y., \& {Heger}, A. 2021, \aap, 654,
  A103, \dodoi{10.1051/0004-6361/202141518}

\bibitem[{{Liu} \& {Stancliffe}(2018)}]{Liu2018}
{Liu}, Z.-W., \& {Stancliffe}, R.~J. 2018, \mnras, 475, 5257,
  \dodoi{10.1093/mnras/sty172}

\bibitem[{{Liu} {et~al.}(2015){Liu}, {Stancliffe}, {Abate}, \&
  {Wang}}]{Liu2015}
{Liu}, Z.-W., {Stancliffe}, R.~J., {Abate}, C., \& {Wang}, B. 2015, \apj, 808,
  138, \dodoi{10.1088/0004-637X/808/2/138}

\bibitem[{{Liu} \& {Zeng}(2021)}]{Liu2021a}
{Liu}, Z.-W., \& {Zeng}, Y. 2021, \mnras, 500, 301,
  \dodoi{10.1093/mnras/staa3280}

\bibitem[{{Liu} {et~al.}(2013{\natexlab{c}}){Liu}, {Pakmor}, {Seitenzahl},
  {Hillebrandt}, {Kromer}, {R{\"o}pke}, {Edelmann}, {Taubenberger}, {Maeda},
  {Wang}, \& {Han}}]{Liu2013b}
{Liu}, Z.-W., {Pakmor}, R., {Seitenzahl}, I.~R., {et~al.} 2013{\natexlab{c}},
  \apj, 774, 37, \dodoi{10.1088/0004-637X/774/1/37}

\bibitem[{{Livio} \& {Riess}(2003)}]{Livio2003}
{Livio}, M., \& {Riess}, A.~G. 2003, \apjl, 594, L93, \dodoi{10.1086/378765}

\bibitem[{{Lundqvist} {et~al.}(2015){Lundqvist}, {Nyholm}, {Taddia},
  {Sollerman}, {Johansson}, {Kozma}, {Lundqvist}, {Fransson}, {Garnavich},
  {Kromer}, {Shappee}, \& {Goobar}}]{Lundqvist2015}
{Lundqvist}, P., {Nyholm}, A., {Taddia}, F., {et~al.} 2015, \aap, 577, A39,
  \dodoi{10.1051/0004-6361/201525719}

\bibitem[{{Maoz} \& {Mannucci}(2008)}]{Maoz2008}
{Maoz}, D., \& {Mannucci}, F. 2008, \mnras, 388, 421,
  \dodoi{10.1111/j.1365-2966.2008.13403.x}

\bibitem[{{Maoz} {et~al.}(2014){Maoz}, {Mannucci}, \& {Nelemans}}]{Maoz2014}
{Maoz}, D., {Mannucci}, F., \& {Nelemans}, G. 2014, \araa, 52, 107,
  \dodoi{10.1146/annurev-astro-082812-141031}

\bibitem[{{Margutti} {et~al.}(2012){Margutti}, {Soderberg}, {Chomiuk},
  {Chevalier}, {Hurley}, {Milisavljevic}, {Foley}, {Hughes}, {Slane},
  {Fransson}, {Moe}, {Barthelmy}, {Boynton}, {Briggs}, {Connaughton}, {Costa},
  {Cummings}, {Del Monte}, {Enos}, {Fellows}, {Feroci}, {Fukazawa}, {Gehrels},
  {Goldsten}, {Golovin}, {Hanabata}, {Harshman}, {Krimm}, {Litvak},
  {Makishima}, {Marisaldi}, {Mitrofanov}, {Murakami}, {Ohno}, {Palmer},
  {Sanin}, {Starr}, {Svinkin}, {Takahashi}, {Tashiro}, {Terada}, \&
  {Yamaoka}}]{Margutti2012}
{Margutti}, R., {Soderberg}, A.~M., {Chomiuk}, L., {et~al.} 2012, \apj, 751,
  134, \dodoi{10.1088/0004-637X/751/2/134}

\bibitem[{{Marietta} {et~al.}(2000){Marietta}, {Burrows}, \&
  {Fryxell}}]{Marietta2000}
{Marietta}, E., {Burrows}, A., \& {Fryxell}, B. 2000, \apjs, 128, 615,
  \dodoi{10.1086/313392}

\bibitem[{{McCully} {et~al.}(2014){McCully}, {Jha}, {Foley}, {Bildsten},
  {Fong}, {Kirshner}, {Marion}, {Riess}, \& {Stritzinger}}]{McCully2014}
{McCully}, C., {Jha}, S.~W., {Foley}, R.~J., {et~al.} 2014, \nat, 512, 54,
  \dodoi{10.1038/nature13615}

\bibitem[{{McCully} {et~al.}(2021){McCully}, {Jha}, {Scalzo}, {Howell},
  {Foley}, {Zeng}, {Liu}, {Hosseinzadeh}, {Bildsten}, {Riess}, {Kirshner},
  {Marion}, \& {Camacho-Neves}}]{McCully2021}
{McCully}, C., {Jha}, S.~W., {Scalzo}, R.~A., {et~al.} 2021, arXiv e-prints,
  arXiv:2106.04602.
\newblock \doarXiv{2106.04602}

\bibitem[{{Nomoto}(1982)}]{Nomoto1982}
{Nomoto}, K. 1982, \apj, 253, 798, \dodoi{10.1086/159682}

\bibitem[{{Nomoto} {et~al.}(1984){Nomoto}, {Thielemann}, \&
  {Yokoi}}]{Nomoto1984}
{Nomoto}, K., {Thielemann}, F.-K., \& {Yokoi}, K. 1984, \apj, 286, 644,
  \dodoi{10.1086/162639}

\bibitem[{{Nugent} {et~al.}(2011){Nugent}, {Sullivan}, {Cenko}, {Thomas},
  {Kasen}, {Howell}, {Bersier}, {Bloom}, {Kulkarni}, {Kandrashoff},
  {Filippenko}, {Silverman}, {Marcy}, {Howard}, {Isaacson}, {Maguire},
  {Suzuki}, {Tarlton}, {Pan}, {Bildsten}, {Fulton}, {Parrent}, {Sand},
  {Podsiadlowski}, {Bianco}, {Dilday}, {Graham}, {Lyman}, {James}, {Kasliwal},
  {Law}, {Quimby}, {Hook}, {Walker}, {Mazzali}, {Pian}, {Ofek}, {Gal-Yam}, \&
  {Poznanski}}]{Nugent2011}
{Nugent}, P.~E., {Sullivan}, M., {Cenko}, S.~B., {et~al.} 2011, \nat, 480, 344,
  \dodoi{10.1038/nature10644}

\bibitem[{{Pagnotta} \& {Schaefer}(2015)}]{Pagnotta2015}
{Pagnotta}, A., \& {Schaefer}, B.~E. 2015, \apj, 799, 101,
  \dodoi{10.1088/0004-637X/799/1/101}

\bibitem[{{Pagnotta} {et~al.}(2014){Pagnotta}, {Walker}, \&
  {Schaefer}}]{Pagnotta2014}
{Pagnotta}, A., {Walker}, E.~S., \& {Schaefer}, B.~E. 2014, \apj, 788, 173,
  \dodoi{10.1088/0004-637X/788/2/173}

\bibitem[{{Pakmor} {et~al.}(2012){Pakmor}, {Edelmann}, {R{\"o}pke}, \&
  {Hillebrandt}}]{Pakmor2012a}
{Pakmor}, R., {Edelmann}, P., {R{\"o}pke}, F.~K., \& {Hillebrandt}, W. 2012,
  \mnras, 424, 2222, \dodoi{10.1111/j.1365-2966.2012.21383.x}

\bibitem[{{Pakmor} {et~al.}(2010){Pakmor}, {Kromer}, {R{\"o}pke}, {Sim},
  {Ruiter}, \& {Hillebrandt}}]{Pakmor2010}
{Pakmor}, R., {Kromer}, M., {R{\"o}pke}, F.~K., {et~al.} 2010, \nat, 463, 61,
  \dodoi{10.1038/nature08642}

\bibitem[{{Pakmor} {et~al.}(2008){Pakmor}, {R{\"o}pke}, {Weiss}, \&
  {Hillebrandt}}]{Pakmor2008}
{Pakmor}, R., {R{\"o}pke}, F.~K., {Weiss}, A., \& {Hillebrandt}, W. 2008, \aap,
  489, 943, \dodoi{10.1051/0004-6361:200810456}

\bibitem[{{Pan} {et~al.}(2012{\natexlab{a}}){Pan}, {Ricker}, \&
  {Taam}}]{Pan2012a}
{Pan}, K.-C., {Ricker}, P.~M., \& {Taam}, R.~E. 2012{\natexlab{a}}, \apj, 750,
  151, \dodoi{10.1088/0004-637X/750/2/151}

\bibitem[{{Pan} {et~al.}(2012{\natexlab{b}}){Pan}, {Ricker}, \&
  {Taam}}]{Pan2012b}
---. 2012{\natexlab{b}}, \apj, 760, 21, \dodoi{10.1088/0004-637X/760/1/21}

\bibitem[{{Pan} {et~al.}(2013){Pan}, {Ricker}, \& {Taam}}]{Pan2013}
---. 2013, \apj, 773, 49, \dodoi{10.1088/0004-637X/773/1/49}

\bibitem[{{Paxton} {et~al.}(2011){Paxton}, {Bildsten}, {Dotter}, {Herwig},
  {Lesaffre}, \& {Timmes}}]{Paxton2011}
{Paxton}, B., {Bildsten}, L., {Dotter}, A., {et~al.} 2011, \apjs, 192, 3,
  \dodoi{10.1088/0067-0049/192/1/3}

\bibitem[{{Paxton} {et~al.}(2013){Paxton}, {Cantiello}, {Arras}, {Bildsten},
  {Brown}, {Dotter}, {Mankovich}, {Montgomery}, {Stello}, {Timmes}, \&
  {Townsend}}]{Paxton2013}
{Paxton}, B., {Cantiello}, M., {Arras}, P., {et~al.} 2013, \apjs, 208, 4,
  \dodoi{10.1088/0067-0049/208/1/4}

\bibitem[{{Paxton} {et~al.}(2015){Paxton}, {Marchant}, {Schwab}, {Bauer},
  {Bildsten}, {Cantiello}, {Dessart}, {Farmer}, {Hu}, {Langer}, {Townsend},
  {Townsley}, \& {Timmes}}]{Paxton2015}
{Paxton}, B., {Marchant}, P., {Schwab}, J., {et~al.} 2015, \apjs, 220, 15,
  \dodoi{10.1088/0067-0049/220/1/15}

\bibitem[{{Paxton} {et~al.}(2018){Paxton}, {Schwab}, {Bauer}, {Bildsten},
  {Blinnikov}, {Duffell}, {Farmer}, {Goldberg}, {Marchant}, {Sorokina},
  {Thoul}, {Townsend}, \& {Timmes}}]{Paxton2018}
{Paxton}, B., {Schwab}, J., {Bauer}, E.~B., {et~al.} 2018, \apjs, 234, 34,
  \dodoi{10.3847/1538-4365/aaa5a8}

\bibitem[{{Perlmutter} {et~al.}(1999){Perlmutter}, {Aldering}, {Goldhaber},
  {Knop}, {Nugent}, {Castro}, {Deustua}, {Fabbro}, {Goobar}, {Groom}, {Hook},
  {Kim}, {Kim}, {Lee}, {Nunes}, {Pain}, {Pennypacker}, {Quimby}, {Lidman},
  {Ellis}, {Irwin}, {McMahon}, {Ruiz-Lapuente}, {Walton}, {Schaefer}, {Boyle},
  {Filippenko}, {Matheson}, {Fruchter}, {Panagia}, {Newberg}, {Couch}, \& {The
  Supernova Cosmology Project}}]{Perlmutter1999}
{Perlmutter}, S., {Aldering}, G., {Goldhaber}, G., {et~al.} 1999, \apj, 517,
  565, \dodoi{10.1086/307221}

\bibitem[{{Phillips}(1993)}]{Phillips1993}
{Phillips}, M.~M. 1993, \apjl, 413, L105, \dodoi{10.1086/186970}

\bibitem[{{Phillips} {et~al.}(1999){Phillips}, {Lira}, {Suntzeff}, {Schommer},
  {Hamuy}, \& {Maza}}]{Phillips1999}
{Phillips}, M.~M., {Lira}, P., {Suntzeff}, N.~B., {et~al.} 1999, \aj, 118,
  1766, \dodoi{10.1086/301032}

\bibitem[{{Podsiadlowski}(2003)}]{Podsiadlowski2003}
{Podsiadlowski}, P. 2003, arXiv e-prints, astro.
\newblock \doarXiv{astro-ph/0303660}

\bibitem[{{Rauscher} {et~al.}(2002){Rauscher}, {Heger}, {Hoffman}, \&
  {Woosley}}]{Rauscher2002}
{Rauscher}, T., {Heger}, A., {Hoffman}, R.~D., \& {Woosley}, S.~E. 2002, \apj,
  576, 323, \dodoi{10.1086/341728}

\bibitem[{{Riess} {et~al.}(1998){Riess}, {Filippenko}, {Challis},
  {Clocchiatti}, {Diercks}, {Garnavich}, {Gilliland}, {Hogan}, {Jha},
  {Kirshner}, {Leibundgut}, {Phillips}, {Reiss}, {Schmidt}, {Schommer},
  {Smith}, {Spyromilio}, {Stubbs}, {Suntzeff}, \& {Tonry}}]{Riess1998}
{Riess}, A.~G., {Filippenko}, A.~V., {Challis}, P., {et~al.} 1998, \aj, 116,
  1009, \dodoi{10.1086/300499}

\bibitem[{{R{\"o}pke} {et~al.}(2007){R{\"o}pke}, {Hillebrandt}, {Schmidt},
  {Niemeyer}, {Blinnikov}, \& {Mazzali}}]{Roepke2007a}
{R{\"o}pke}, F.~K., {Hillebrandt}, W., {Schmidt}, W., {et~al.} 2007, \apj, 668,
  1132, \dodoi{10.1086/521347}

\bibitem[{{R{\"o}pke} \& {Niemeyer}(2007)}]{Roepke2007b}
{R{\"o}pke}, F.~K., \& {Niemeyer}, J.~C. 2007, \aap, 464, 683,
  \dodoi{10.1051/0004-6361:20066585}

\bibitem[{{R{\"o}pke} {et~al.}(2012){R{\"o}pke}, {Kromer}, {Seitenzahl},
  {Pakmor}, {Sim}, {Taubenberger}, {Ciaraldi-Schoolmann}, {Hillebrandt},
  {Aldering}, {Antilogus}, {Baltay}, {Benitez-Herrera}, {Bongard}, {Buton},
  {Canto}, {Cellier-Holzem}, {Childress}, {Chotard}, {Copin}, {Fakhouri},
  {Fink}, {Fouchez}, {Gangler}, {Guy}, {Hachinger}, {Hsiao}, {Chen},
  {Kerschhaggl}, {Kowalski}, {Nugent}, {Paech}, {Pain}, {Pecontal}, {Pereira},
  {Perlmutter}, {Rabinowitz}, {Rigault}, {Runge}, {Saunders}, {Smadja},
  {Suzuki}, {Tao}, {Thomas}, {Tilquin}, \& {Wu}}]{Roepke2012}
{R{\"o}pke}, F.~K., {Kromer}, M., {Seitenzahl}, I.~R., {et~al.} 2012, \apjl,
  750, L19, \dodoi{10.1088/2041-8205/750/1/L19}

\bibitem[{{Rosswog} {et~al.}(2009){Rosswog}, {Kasen}, {Guillochon}, \&
  {Ramirez-Ruiz}}]{Rosswog2009}
{Rosswog}, S., {Kasen}, D., {Guillochon}, J., \& {Ramirez-Ruiz}, E. 2009,
  \apjl, 705, L128, \dodoi{10.1088/0004-637X/705/2/L128}

\bibitem[{{Ruiz-Lapuente}(2019)}]{Ruiz-Lapuente2019}
{Ruiz-Lapuente}, P. 2019, \nar, 85, 101523, \dodoi{10.1016/j.newar.2019.101523}

\bibitem[{{Ruiz-Lapuente} {et~al.}(2018){Ruiz-Lapuente}, {Damiani}, {Bedin},
  {Gonz{\'a}lez Hern{\'a}ndez}, {Galbany}, {Pritchard}, {Canal}, \&
  {M{\'e}ndez}}]{Ruiz-Lapuente2018}
{Ruiz-Lapuente}, P., {Damiani}, F., {Bedin}, L., {et~al.} 2018, \apj, 862, 124,
  \dodoi{10.3847/1538-4357/aac9c4}

\bibitem[{{Ruiz-Lapuente} {et~al.}(2004){Ruiz-Lapuente}, {Comeron},
  {M{\'e}ndez}, {Canal}, {Smartt}, {Filippenko}, {Kurucz}, {Chornock}, {Foley},
  {Stanishev}, \& {Ibata}}]{Ruiz-Lapuente2004}
{Ruiz-Lapuente}, P., {Comeron}, F., {M{\'e}ndez}, J., {et~al.} 2004, \nat, 431,
  1069, \dodoi{10.1038/nature03006}

\bibitem[{{Schaefer} \& {Pagnotta}(2012)}]{Schaefer2012}
{Schaefer}, B.~E., \& {Pagnotta}, A. 2012, \nat, 481, 164,
  \dodoi{10.1038/nature10692}

\bibitem[{{Schmidt} {et~al.}(1998){Schmidt}, {Suntzeff}, {Phillips},
  {Schommer}, {Clocchiatti}, {Kirshner}, {Garnavich}, {Challis}, {Leibundgut},
  {Spyromilio}, {Riess}, {Filippenko}, {Hamuy}, {Smith}, {Hogan}, {Stubbs},
  {Diercks}, {Reiss}, {Gilliland}, {Tonry}, {Maza}, {Dressler}, {Walsh}, \&
  {Ciardullo}}]{Schmidt1998}
{Schmidt}, B.~P., {Suntzeff}, N.~B., {Phillips}, M.~M., {et~al.} 1998, \apj,
  507, 46, \dodoi{10.1086/306308}

\bibitem[{{Seitenzahl} {et~al.}(2013){Seitenzahl}, {Ciaraldi-Schoolmann},
  {R{\"o}pke}, {Fink}, {Hillebrandt}, {Kromer}, {Pakmor}, {Ruiter}, {Sim}, \&
  {Taubenberger}}]{Seitenzahl2013}
{Seitenzahl}, I.~R., {Ciaraldi-Schoolmann}, F., {R{\"o}pke}, F.~K., {et~al.}
  2013, \mnras, 429, 1156, \dodoi{10.1093/mnras/sts402}

\bibitem[{{Shappee} {et~al.}(2013){Shappee}, {Kochanek}, \&
  {Stanek}}]{Shappee2013}
{Shappee}, B.~J., {Kochanek}, C.~S., \& {Stanek}, K.~Z. 2013, \apj, 765, 150,
  \dodoi{10.1088/0004-637X/765/2/150}

\bibitem[{{Shappee} \& {Stanek}(2011)}]{Shappee2011}
{Shappee}, B.~J., \& {Stanek}, K.~Z. 2011, \apj, 733, 124,
  \dodoi{10.1088/0004-637X/733/2/124}

\bibitem[{{Shappee} {et~al.}(2017){Shappee}, {Stanek}, {Kochanek}, \&
  {Garnavich}}]{Shappee2017}
{Shappee}, B.~J., {Stanek}, K.~Z., {Kochanek}, C.~S., \& {Garnavich}, P.~M.
  2017, \apj, 841, 48, \dodoi{10.3847/1538-4357/aa6eab}

\bibitem[{{Shen} \& {Bildsten}(2007)}]{Shen2007}
{Shen}, K.~J., \& {Bildsten}, L. 2007, \apj, 660, 1444, \dodoi{10.1086/513457}

\bibitem[{{Shen} {et~al.}(2018){Shen}, {Boubert}, {G{\"a}nsicke}, {Jha},
  {Andrews}, {Chomiuk}, {Foley}, {Fraser}, {Gromadzki}, {Guillochon}, {Kotze},
  {Maguire}, {Siebert}, {Smith}, {Strader}, {Badenes}, {Kerzendorf}, {Koester},
  {Kromer}, {Miles}, {Pakmor}, {Schwab}, {Toloza}, {Toonen}, {Townsley}, \&
  {Williams}}]{Shen2018}
{Shen}, K.~J., {Boubert}, D., {G{\"a}nsicke}, B.~T., {et~al.} 2018, \apj, 865,
  15, \dodoi{10.3847/1538-4357/aad55b}

\bibitem[{{Sim} {et~al.}(2010){Sim}, {R{\"o}pke}, {Hillebrandt}, {Kromer},
  {Pakmor}, {Fink}, {Ruiter}, \& {Seitenzahl}}]{Sim2010}
{Sim}, S.~A., {R{\"o}pke}, F.~K., {Hillebrandt}, W., {et~al.} 2010, \apjl, 714,
  L52, \dodoi{10.1088/2041-8205/714/1/L52}

\bibitem[{{Soker}(2019)}]{Soker2019}
{Soker}, N. 2019, \nar, 87, 101535, \dodoi{10.1016/j.newar.2020.101535}

\bibitem[{{Springel}(2005)}]{Springel2005}
{Springel}, V. 2005, \mnras, 364, 1105,
  \dodoi{10.1111/j.1365-2966.2005.09655.x}

\bibitem[{{Springel} {et~al.}(2001){Springel}, {Yoshida}, \&
  {White}}]{Springel2001}
{Springel}, V., {Yoshida}, N., \& {White}, S.~D.~M. 2001, \na, 6, 79,
  \dodoi{10.1016/S1384-1076(01)00042-2}

\bibitem[{{Townsley} {et~al.}(2019){Townsley}, {Miles}, {Shen}, \&
  {Kasen}}]{Townsley2019}
{Townsley}, D.~M., {Miles}, B.~J., {Shen}, K.~J., \& {Kasen}, D. 2019, \apjl,
  878, L38, \dodoi{10.3847/2041-8213/ab27cd}

\bibitem[{{Tucker} {et~al.}(2021{\natexlab{a}}){Tucker}, {Ashall}, {Shappee},
  {Kochanek}, {Stanek}, \& {Garnavich}}]{Tucker2021a}
{Tucker}, M.~A., {Ashall}, C., {Shappee}, B.~J., {et~al.} 2021{\natexlab{a}},
  arXiv e-prints, arXiv:2111.00016.
\newblock \doarXiv{2111.00016}

\bibitem[{{Tucker} {et~al.}(2021{\natexlab{b}}){Tucker}, {Shappee}, {Kochanek},
  {Stanek}, {Ashall}, {Anand}, \& {Garnavich}}]{Tucker2021b}
{Tucker}, M.~A., {Shappee}, B.~J., {Kochanek}, C.~S., {et~al.}
  2021{\natexlab{b}}, arXiv e-prints, arXiv:2111.01144.
\newblock \doarXiv{2111.01144}

\bibitem[{{Wang} {et~al.}(2009){Wang}, {Chen}, {Meng}, \& {Han}}]{Wang2009}
{Wang}, B., {Chen}, X., {Meng}, X., \& {Han}, Z. 2009, \apj, 701, 1540,
  \dodoi{10.1088/0004-637X/701/2/1540}

\bibitem[{{Weaver} {et~al.}(1978){Weaver}, {Zimmerman}, \&
  {Woosley}}]{Weaver1978}
{Weaver}, T.~A., {Zimmerman}, G.~B., \& {Woosley}, S.~E. 1978, \apj, 225, 1021,
  \dodoi{10.1086/156569}

\bibitem[{{Webbink}(1984)}]{Webbink1984}
{Webbink}, R.~F. 1984, \apj, 277, 355, \dodoi{10.1086/161701}

\bibitem[{{Wheeler} {et~al.}(1975){Wheeler}, {Lecar}, \& {McKee}}]{Wheeler1975}
{Wheeler}, J.~C., {Lecar}, M., \& {McKee}, C.~F. 1975, \apj, 200, 145,
  \dodoi{10.1086/153771}

\bibitem[{{Whelan} \& {Iben}(1973)}]{Whelan1973}
{Whelan}, J., \& {Iben}, Jr., I. 1973, \apj, 186, 1007, \dodoi{10.1086/152565}

\bibitem[{{Woosley} {et~al.}(2002){Woosley}, {Heger}, \&
  {Weaver}}]{Woosley2002}
{Woosley}, S.~E., {Heger}, A., \& {Weaver}, T.~A. 2002, Reviews of Modern
  Physics, 74, 1015, \dodoi{10.1103/RevModPhys.74.1015}

\bibitem[{{Woosley} {et~al.}(1986){Woosley}, {Taam}, \& {Weaver}}]{Woosley1986}
{Woosley}, S.~E., {Taam}, R.~E., \& {Weaver}, T.~A. 1986, \apj, 301, 601,
  \dodoi{10.1086/163926}

\bibitem[{{Zeng} {et~al.}(2020){Zeng}, {Liu}, \& {Han}}]{Zeng2020}
{Zeng}, Y., {Liu}, Z.-W., \& {Han}, Z. 2020, \apj, 898, 12,
  \dodoi{10.3847/1538-4357/ab9943}

\end{thebibliography}
\bibliographystyle{aasjournal}

\end{document}